\documentclass[referee,pdflatex,template/sn-nature]{template/sn-jnl}

\usepackage[mathlines]{lineno}
\usepackage{graphicx}%
\usepackage{multirow}%
\usepackage{amsmath,amssymb,amsfonts}%
\usepackage{amsthm}%
\usepackage{mathrsfs}%
\usepackage[title]{appendix}%
\usepackage{xcolor}%
\usepackage{textcomp}%
\usepackage{manyfoot}%
\usepackage{booktabs}%
\usepackage{algorithm}%
\usepackage{algorithmicx}%
\usepackage{algpseudocode}%
\usepackage{listings}%
\usepackage[super,sort&compress]{natbib}%
\usepackage{bbm}
\usepackage[acronym,nonumberlist,nopostdot]{glossaries}
\makeglossaries

\usepackage{pdflscape} 
\usepackage{tabularx,siunitx}
\newcolumntype{L}{>{\raggedright\arraybackslash}X}
\newacronym{ai}{AI}{artificial intelligence}
\newacronym{ml}{ML}{machine learning}
\newacronym{fst}{FST}{Fitzpatrick skin type}
\newacronym{ssl}{SSL}{self-supervised learning}
\newacronym{fd}{FD}{Fr\'echet distance}
\newacronym{tbp}{TBP}{total body photography}
\newacronym{llm}{LLM}{large language model}
\newacronym{icd}{ICD}{international classification of diseases}

\theoremstyle{thmstyleone}%
\theoremstyle{thmstyletwo}%

\theoremstyle{thmstylethree}%

\begin{document}
\frenchspacing

\title[SkinMap]{
\centering
A Global Atlas of Digital Dermatology \\[-0.5em]
to Map Innovation and Disparities
}

\author*[1,2]{\fnm{Fabian} \sur{Gröger}}\email{fabian.groeger@unibas.ch}
\author[2]{\fnm{Simone} \sur{Lionetti}}\email{simone.lionetti@hslu.ch}
\author[1,3]{\fnm{Philippe} \sur{Gottfrois}}\email{philippe.gottfrois@unibas.ch}
\author[2,3]{\fnm{Alvaro} \sur{Gonzalez-Jimenez}}\email{alvaro.gonzalezjimenez@usb.ch}
\author[3]{\fnm{Lea} \sur{Habermacher}}\email{lea.habermacher@usb.ch}
\author[3]{\fnm{Labelling} \sur{Consortium}}
\consortiumcont{Labelling Consortium: Maximilian Boehm, Sara Cerminara, Georgios Nikolakis, and Alexander Romswinkel.}
\author[2]{\fnm{Ludovic} \sur{Amruthalingam}}\email{ludovic.amruthalingam@hslu.ch}
\author[4]{\fnm{Matthew} \sur{Groh}}\email{matthew.groh@kellogg.northwestern.edu}
\author[2]{\fnm{Marc} \sur{Pouly}}\email{marc.pouly@hslu.ch}
\equalcont{These authors contributed equally to this work.}
\author*[1,3]{\fnm{Alexander A.} \sur{Navarini}}\email{alexander.navarini@usb.ch}
\equalcont{These authors contributed equally to this work.}

\affil[1]{\orgdiv{Department of Biomedical Engineering}, \orgname{University of Basel}, \orgaddress{\street{Hegenheimermattweg 167b}, \city{Allschwil}, \postcode{CH--4123}, \state{Basel}, \country{Switzerland}}}
\affil[2]{\orgdiv{Department of Computer Science}, \orgname{Lucerne University of Applied Sciences and Arts}, \orgaddress{\street{Suurstoffi 1}, \city{Rotkreuz}, \postcode{CH--6343}, \state{Zug}, \country{Switzerland}}}
\affil[3]{\orgdiv{Department of Dermatology}, \orgname{University Hospital of Basel}, \orgaddress{\street{Burgfelderstrasse 101}, \city{Basel}, \postcode{CH--4055}, \state{Basel}, \country{Switzerland}}}
\affil[4]{\orgdiv{Kellogg School of Management}, \orgname{Northwestern University}, \orgaddress{\street{2211 Campus Drive}, \city{Evanston}, \postcode{US--60208}, \state{Illinois}, \country{United States of America}}}

\abstract{
\looseness=-1
The adoption of artificial intelligence in dermatology promises democratized access to healthcare, but model reliability depends on the quality and comprehensiveness of the data fueling these models.
Despite rapid growth in publicly available dermatology images, the field lacks quantitative key performance indicators to measure whether new datasets expand clinical coverage or merely replicate what is already known. 
Here we present \textsc{SkinMap}, a multi-modal framework for the first comprehensive audit of the field's entire data basis.
We unify the publicly available dermatology datasets into a single, queryable semantic atlas comprising more than 1.1~million images of skin conditions and quantify (i) informational novelty over time, (ii) dataset redundancy, and (iii) representation gaps across demographics and diagnoses.
Despite exponential growth in dataset sizes, informational novelty across time has somewhat plateaued: Some clusters, such as common neoplasms on fair skin, are densely populated, while underrepresented skin types and many rare diseases remain unaddressed.
We further identify structural gaps in coverage: Darker skin tones (Fitzpatrick V--VI) constitute only 5.8\% of images and pediatric patients only 3.0\%, while many rare diseases and phenotype combinations remain sparsely represented. 
\textsc{SkinMap} provides infrastructure to measure blind spots and steer strategic data acquisition toward undercovered regions of clinical space.
}

\keywords{Artificial intelligence, Dermatology, Health equity, Data diversity} 
\maketitle

\begin{figure}[htbp]
    \centering
    \includegraphics[width=1.0\linewidth]{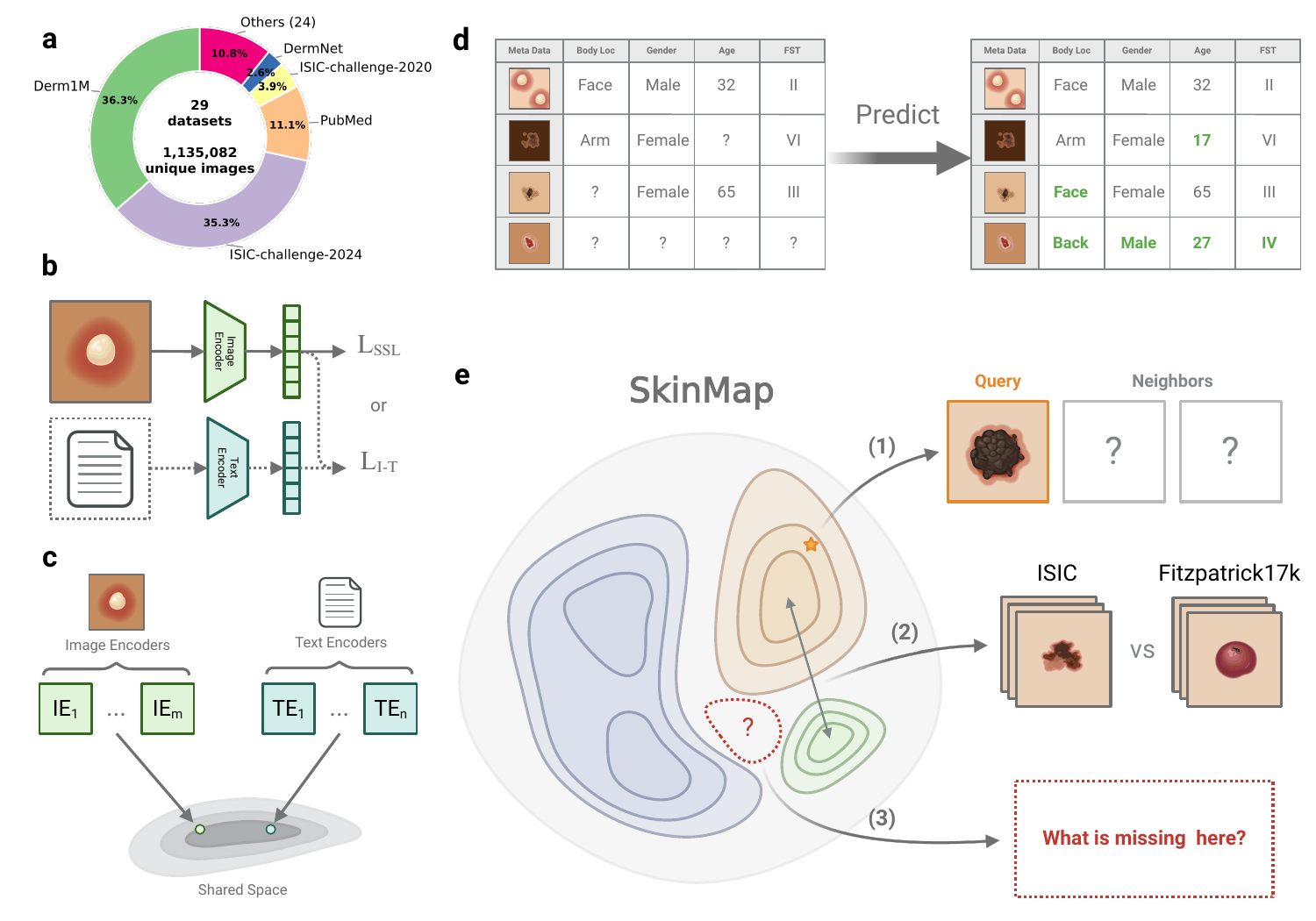}
    \caption{
    \textbf{The \textsc{SkinMap} multi-modal framework for auditing the global dermatology landscape.} 
    \textbf{a}, Aggregation of 29 public datasets yielding 1.1 million unique images with sparse original metadata. 
    \textbf{b}, Multi-modal training pipeline utilizing \gls{ssl} on imaging data and contrastive learning on image-text pairs generated via templated captions. 
    \textbf{c}, Construction of a unified latent space by projecting diverse encoders into a shared low-dimensional manifold. 
    \textbf{d}, Imputation engine training, where linear probes learn to predict missing attributes (\textit{e.g.}, \acrlong{fst}, age, and sex) using partial meta-information. 
    \textbf{e}, Applications of the resulting digital \textsc{SkinMap} atlas: 
    (1) A clinician-facing search tool for finding semantically similar cases;
    (2) a structural audit tool for identifying dataset overlaps and homologies;
    and (3) a strategic acquisition guide for detecting underrepresented latent regions to direct future data collection.
    }
    \label{fig:Method}
\end{figure}

\section*{Introduction}\label{sec:introduction}
\looseness=-1
The integration of \gls{ai} into medicine revolutionizes many clinical specialties, including dermatology. 
However, the performance and clinical reliability of these models are fundamentally dependent on the quality, diversity, and scale of the data on which they are trained \citep{esteva2017dermatologist,pmlr-v139-hashimoto21a,goyal2024scaling,subramanyam2025scaling}. 
For years, the dermatology community has pointed to data scarcity, particularly diversity-related issues, as a primary obstacle to progress \citep{daneshjou2021lack,daneshjou_disparities_2022,salinas2024systematic,groger2025review}, but the claim has remained largely unquantified. 
This gap masks deeper issues: Without a systematic audit of the global data landscape, the research community risks investing resources in collecting redundant data, while biases and blind spots go unaddressed. 
Furthermore, models trained on unexamined corpora of data can inadvertently learn and amplify biases related to skin tones, rare conditions, or specific demographics, leading to tools that fail in real-world clinical settings and potentially worsen existing health disparities.
Previous initiatives to improve the data landscape have been largely unguided, focusing on presumed gaps and biases rather than those identified through systematic quantitative analysis \citep{groh_evaluating_2021,daneshjou_disparities_2022,daneshjou_skincon_2023}.
Although these contributions are valuable, they lack a comprehensive perspective because a reproducible and scalable methodology to measure dataset coverage and marginal utility is missing. 
Consequently, the field operates without key performance indicators to quantify whether new data expand the clinical map or reinforce existing blind spots.

Here, we introduce \textsc{SkinMap}, a framework that unifies the fragmented global archive into a queryable, interactive digital atlas.
Visualizing the data landscape in real time enables clinicians and researchers to search for clinically similar cases and to transparently assess data availability across collections.
To achieve this, we construct a shared semantic embedding space that captures clinical similarities across datasets. 
We aggregate 1,135,082 unique images from 29 public datasets, ranging from leading archives such as \texttt{ISIC}~\citep{isic2022} to specialized cohorts~\citep{groh_evaluating_2021,10.1001/jamanetworkopen.2024.46615,gottfrois2024passion,madarkar2025dermacon,leibe_benchmark_2016,zhou2024skincap,pacheco_pad-ufes-20_2020,daneshjou_disparities_2022,giotis2015a,tschandl_ham10000_2018,Kawahara2018-7pt,mendonca_ph2_2013,khushbu_akter_2024_skin_disease_classification,mskcc_2025_skin_tone_labeling,ricci2023dataset}, and harmonize them to enable joint analysis. 
By employing a multi-modal approach that aligns images with standardized natural language captions, and combining them in an ensemble with purely image-trained models, we project these disparate samples into a common latent space where proximity corresponds to visual and clinical similarity.
Crucially, this shared space enables us to quantitatively assess the central issue of unharmonized and missing metadata. 
Using the latent representations of annotated samples, we infer missing clinical attributes, such as \gls{fst}, age, gender, and geographic origin, across the entire collection with clinically validated, high-accuracy. 
This imputation transforms the sparsely annotated archive into an actionable resource, enabling us to analyze the demographics of the entire 1.1 million-image collection rather than only the labeled and harmonized subset.
Following unification, our audit indicates that the dermatology data landscape remains unrepresentative of the global population. 
Current datasets are disproportionately composed of lightly pigmented skin images and are geographically concentrated in the Global North~\citep{groh_evaluating_2021,daneshjou2021lack}, whilst darker skin tones (\gls{fst}~V--VI) comprise only 5.8\% of available image data.
Furthermore, a temporal analysis of data acquisition reveals that despite exponential growth in dataset sizes, novelty has plateaued across both visual information and associated labels, revealing a systemic inefficiency in current data collection strategies.
Rather than exploring high-gain gaps in rare conditions and underrepresented demographics, the community is investing scarce resources in the redundant accumulation of common neoplasms among fair-skinned individuals.
Finally, by quantifying how similar datasets truly are, we show that several nominally distinct datasets share near-identical representations, implying that many so-called external validations in the literature may effectively be internal validations.
This unrecognized data leakage artificially inflates performance metrics and risks deploying models that fail to generalize to truly new patient populations.

Ultimately, \textsc{SkinMap} shifts the paradigm from unguided data accumulation to strategic acquisition. 
We delineate high-priority data needs, identify critical gaps such as healthy skin with \gls{fst} V--VI and specific nail pathologies, and provide a concrete, evidence-based road map for guiding future data collection and collaboration.
The \textsc{SkinMap} framework constitutes a generalizable blueprint for other medical imaging domains, such as radiology and pathology, setting a new standard for auditing and expanding the data foundations upon which the next generation of medical \gls{ai} will be built.
To support these research efforts, we release the complete \textsc{SkinMap} framework, including digital atlas and the model ensemble, under an open-source license.

\section*{Results}\label{sec:results}

\subsection*{Quantifying demographic biases}\label{subsec:sample_vs_population}
\begin{figure}[htbp]
    \centering
    \includegraphics[width=1.0\linewidth]{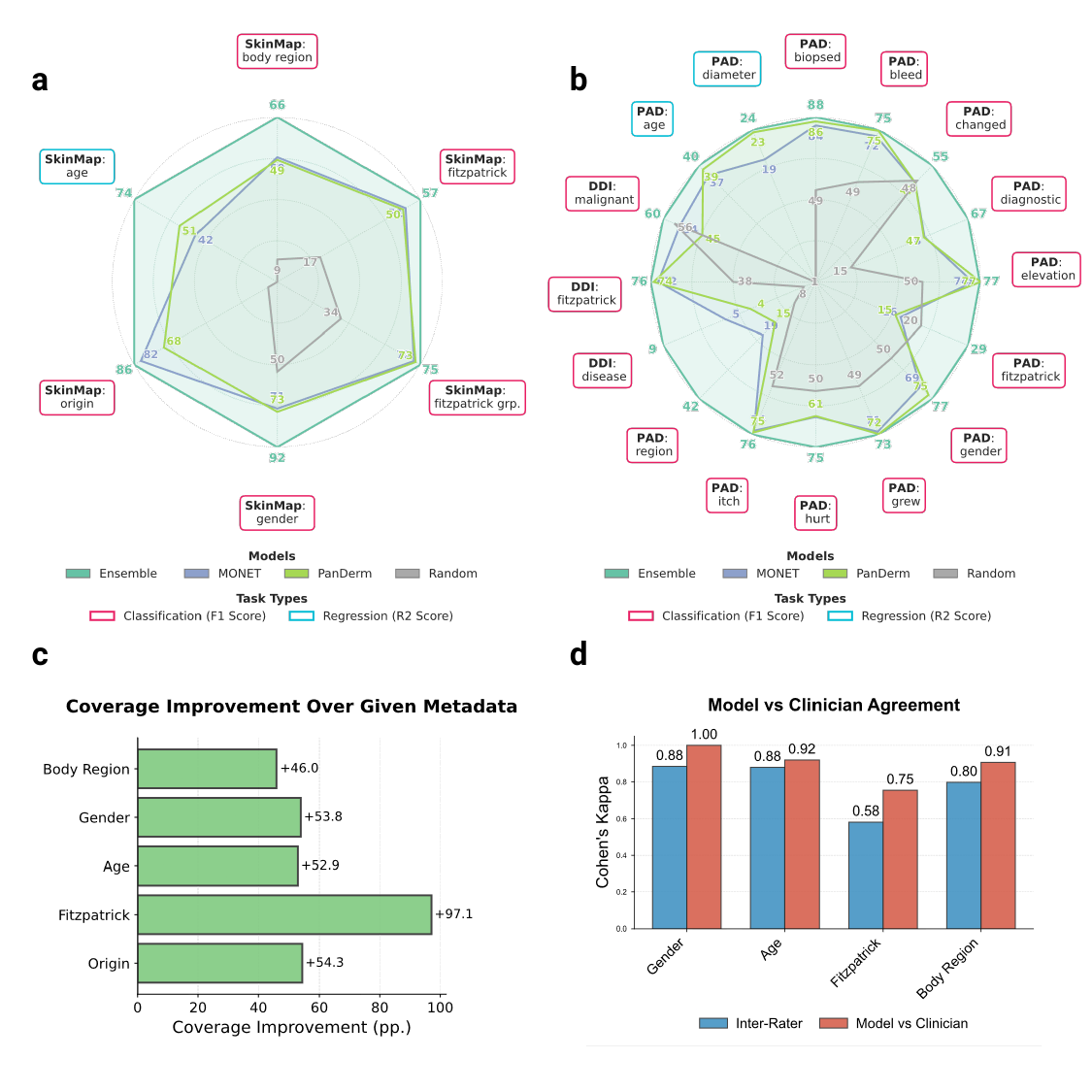}
    \caption{
    \textbf{Benchmarking imputation precision and metadata expansion.} 
    \textbf{a}, \textbf{b}, Radar plots comparing the predictive performance of the \textsc{SkinMap} ensemble against state-of-the-art foundation models MONET and PanDerm across multiple attributes on the internal validation set~(\textbf{a}) and external hold-out datasets~(\textbf{b}) and against random performance. 
    The Ensemble model demonstrates superior performance in recovering missing demographic labels. 
    \textbf{c}, Quantification of metadata coverage expansion achieved via imputation, showing significant gains for key attributes such as \gls{fst} (+97.1 pp.) and geographic origin (+54.3 pp.). 
    \textbf{d}, Comparative analysis of the model's imputation accuracy versus experienced dermatologists on a subset of 150 diverse cases.
    }
    \label{fig:PerformanceComparison}
\end{figure}

\begin{figure}[htbp]
    \centering
    \includegraphics[width=1.0\linewidth]{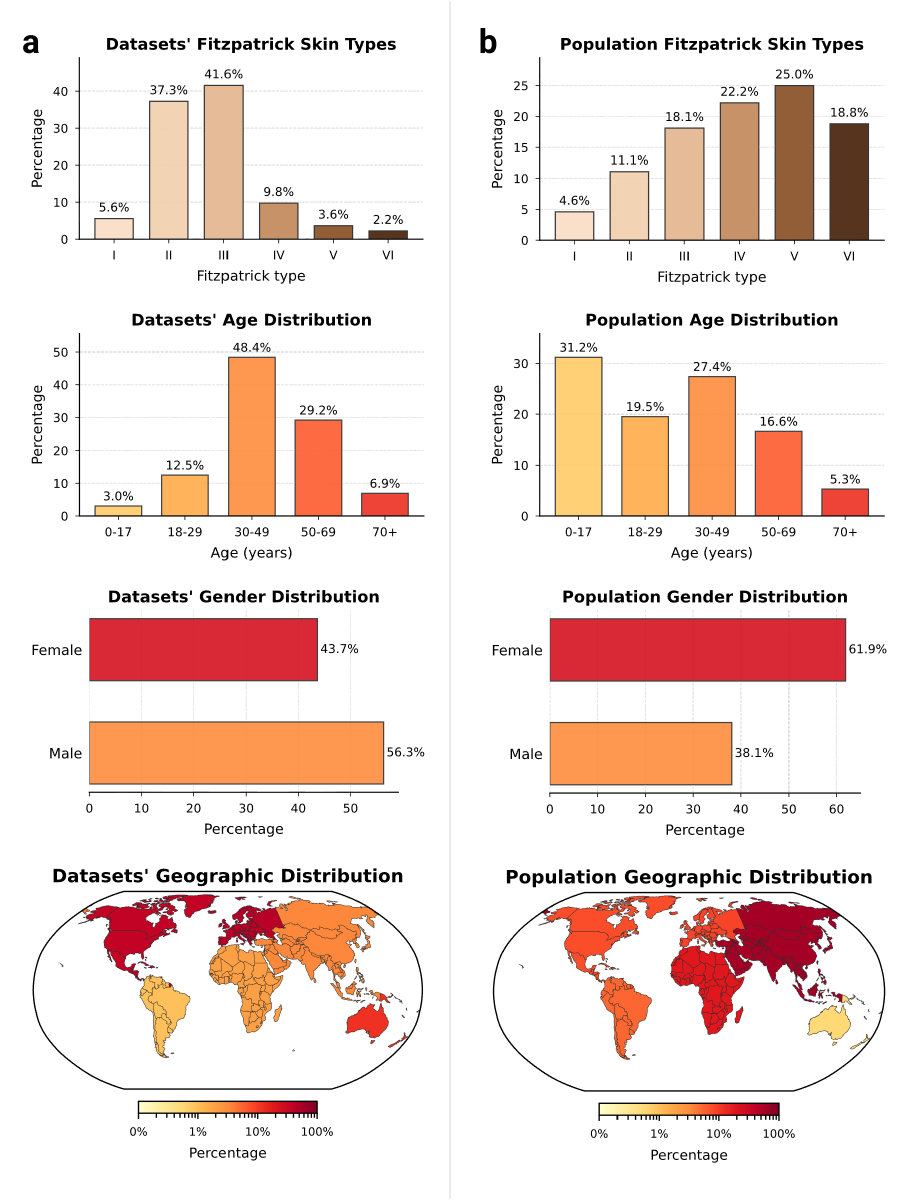}
    \caption{
    \looseness=-1
    \textbf{Systemic demographic and geographic disparities in the global dermatological datasets.}
    Distributions of demographic attributes across the aggregated digital datasets~(left, \textbf{a}) compared with global population statistics for dermatological visits~(right, \textbf{b}). 
    The comparison reveals sharp divergences: \gls{fst} V--VI are marginalized (5.8\% in datasets vs. 43.8\% globally), pediatric populations are underrepresented (3.0\% vs. 31.2\%), and geographic origins are heavily skewed toward the Global North.
    }
    \label{fig:DataRealityCheck}
\end{figure}

The vast majority of the 1.1 million images lack a complete set of demographic metadata, such as skin tone, age, and gender.
To address this, we utilized the \textsc{SkinMap} embedding space to impute missing attributes, expanding the metadata coverage of the dataset by an average of 50.7 percentage points (Fig.~\ref{fig:PerformanceComparison}c). 
We validated this imputation engine on an evaluation set and two datasets (\texttt{DDI}~\citep{daneshjou_disparities_2022} and \texttt{PAD-UFES-20}~\citep{pacheco_pad-ufes-20_2020}) that were strictly held out, \textit{i.e.}, excluded from pretraining. 
The \textsc{SkinMap} ensemble demonstrates superior performance, robustly outperforming state-of-the-art foundation models, including MONET~\citep{kim2024transparent} and PanDerm~\citep{yan2025multimodal}, in attribute prediction accuracy (Fig.~\ref{fig:PerformanceComparison}a--b, Supplementary Table~\ref{tab:model_performance_ci} for statistical confidence intervals).
The strong concordance between the ground-truth subset and the fully imputed archive confirms that the ensemble faithfully estimates the demographics without hallucinating diversity or introducing algorithmic bias (Supplementary Fig.~\ref{fig:DataRealityCheckInclOriginal}).
This is further confirmed by validating the predictions for 150 cases against experienced, practicing dermatologists (n=5), who agree across metadata more with the model than among themselves (Fig.~\ref{fig:PerformanceComparison}d).
This densely annotated dataset overcomes the limitations of the original, sparse labels to assess the demographic composition of the entire digital archive.

Comparing the demographic distribution in the available data with expected population demographics reveals a significant divergence between dataset collection and global demographics (Fig.~\ref{fig:DataRealityCheck}a--b). 
To refute the argument that data scarcity merely reflects workforce shortages, we benchmarked against clinical prevalence statistics, \textit{i.e.}, specifically outpatient encounters~\citep{orenstein2025outpatient} and diverse study cohorts~\citep{he2014self}, alongside WHO global standards~\citep{un2024world,ahmad2001age}. 
The most profound disparity exists in skin tone: While \gls{fst} III--VI comprise about $84.1\%$ of the global dermatology-visit baseline, the available datasets are skewed predominantly toward \gls{fst} II--III ($78.9\%$), with darker skin tones (\gls{fst} V--VI) constituting only $5.8\%$. 
There is an equally strong geographical bias, elsewhere often referred to as a form of ``data colonialism'' \citep{couldry2019costs}, where datasets originate almost exclusively from the Global North (North America, Western Europe) and Australia, leaving the majority of the human population in the Global South virtually unrepresented (Fig.~\ref{fig:DataRealityCheck}c, bottom).
The underrepresented regions of the map highlight an urgent need for targeted, infrastructure-level data collection partnerships with the Global South and across Asia.
We similarly observe significant shifts in age and gender. 
The data collection centers on middle age ($48.4\%$ aged 30--49), with the pediatric population (0--17 years) marginalized at $3.0\%$ compared to $31.2\%$ globally.
This gap likely reflects a combination of true prevalence differences in dominant skin-cancer-focused archives and stricter consent/privacy constraints around pediatric imaging.
Nevertheless, even within general dermatology cohorts, common pediatric conditions (\textit{e.g.}, eczema across diverse skin tones) remain sparsely represented, leaving important portions of pediatric clinical space effectively unmapped.
While gender approaches parity ($43.7\%$ Female), intersectional imbalances compound in specific subgroups, \textit{e.g.}, Female aged 50+ (Supplementary Information~\ref{app:SparseMetadata}).
In summary, the demographic distribution represented by current dermatological data collections is a skewed reflection of the dermatological population: It is predominantly white, western, and middle-aged. 
By quantifying these mismatches, \textsc{SkinMap} moves the field from assuming the existence of bias to precisely knowing where data support is missing, providing a road map for the targeted data acquisition required to build truly global medical \gls{ai}.

\subsection*{Diminishing novelty in data acquisition}\label{subsec:yearly_novelty}

\begin{figure}[H]
    \singlespacing
    \centering
    \includegraphics[width=1.0\linewidth]{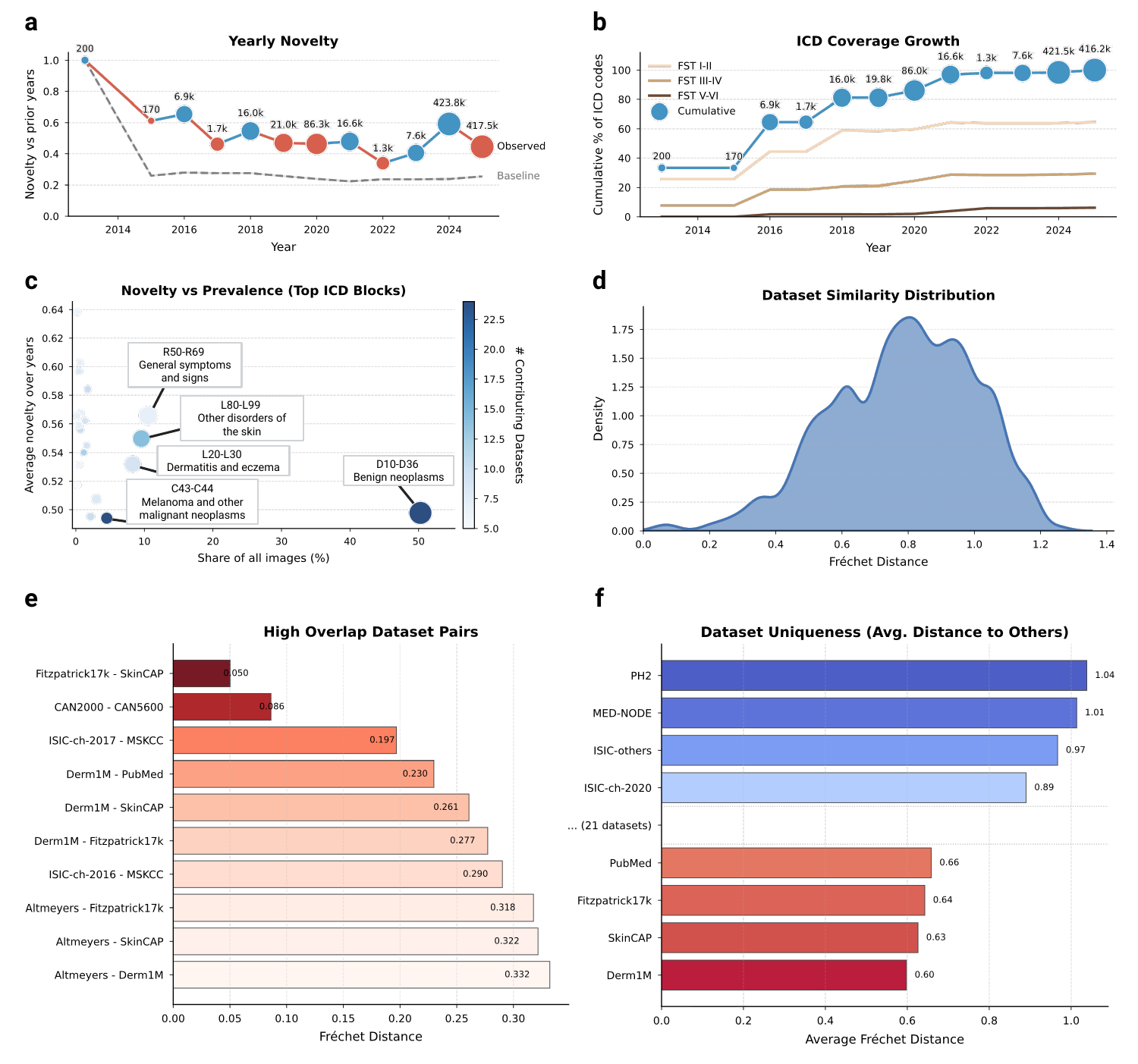}
    \caption{
    \textbf{Diminishing informational novelty and latent dataset redundancy.} 
    \textbf{a}, Temporal analysis of data acquisition showing the yearly novelty (blue or red depending on whether it increases or decreases relative to the previous year), defined as the average embedding distance of new samples relative to the historical pool. 
    Gray shows the random baseline, which represents the expected novelty if new releases were merely random resamples of the historical data pool.
    \textbf{b}, Cumulative percentage of \acrshort{icd}-10 codes covered over time, showing a plateau in diagnostic diversity despite increasing data volume. 
    \textbf{c}, Novelty versus prevalence for top \acrshort{icd} blocks; high-volume categories (\textit{e.g.}, \textit{benign neoplasms}) exhibit low novelty, while high-novelty categories (\textit{e.g.}, \textit{general symptoms}) remain rare. 
    \textbf{d}, Density of pairwise distance between collections, skewed toward low values, indicating high redundancy. 
    \textbf{e}, Identification of ``high overlap'' pairs, revealing hidden subset relationships. 
    \textbf{f}, ``Dataset uniqueness'' scores, highlighting outliers like \texttt{PH2} as high-value sources of distributional variance.
    }
    \label{fig:Novelty_Dataset_Simi}
\end{figure}

Model performance is often assumed to scale with data volume~\citep{kaplan2020scaling}. 
Our temporal audit suggests the field is nearing a saturation regime in which additional samples contribute less and less novel information. 
We quantify novelty as the embedding-space dissimilarity of collected samples relative to previously available data. 
To separate the exploration of new regions from the effect of increasing the sample size of the existing distribution, we compare the collected data against a dataset that is simply resampled from the historical pool.
We define \emph{yearly novelty} as the average distance of the samples released in a given calendar year from their neighbors, divided by the same average for the data resampled from previous years.
Against this baseline, we observe a clear decline: Increases in raw sample counts no longer yield proportional gains in embedding-space novelty (Fig.~\ref{fig:Novelty_Dataset_Simi}a), indicating diminishing marginal informational return under our operational definition.
Despite the large number of new samples collected, exceeding 1,000,000 in 2017--2025, novelty has plateaued near the random baseline. 
The main exception occurs in 2024, when the addition of \gls{tbp} imagery from \texttt{ISIC-challenge-2024} produces a transient spike in novelty.
Together, these results indicate that recent datasets predominantly land within already-populated regions of the data space, and that future gains will require targeted collection in underrepresented regions rather than indiscriminate scaling.

\looseness=-1
To determine if this saturation stems from visual redundancy or a lack of clinical diversity, we analyzed the yearly \gls{icd} coverage growth stratified by skin tone (Fig.~\ref{fig:Novelty_Dataset_Simi}b). 
The analysis reveals a stark divergence: While disease diversity on fair skin (\gls{fst}~I--II) continues to expand, albeit at a reduced rate, coverage for darker skin tones (\gls{fst}~V--VI) has flatlined at near-zero levels. 
This confirms that the field is predominantly including missing diseases for white populations, while the clinical spectrum for the global majority remains unmapped.
By investigating novelty against prevalence in datasets (Fig.~\ref{fig:Novelty_Dataset_Simi}c), we localized this saturation to specific diagnostic categories. 
The current landscape is heavily weighted towards collecting \textit{benign neoplasms} (\mbox{D10--D36}) and \textit{melanoma} (\mbox{C43--C44}), which provide high sample volumes but low novelty scores. 
Conversely, high-novelty categories such as \textit{general symptoms and signs} (\mbox{R50--R69}) remain poorly addressed in terms of both image volume and contributing datasets.

These findings suggest that without a strategic shift towards underrepresented phenotypes, \textit{i.e.}, specifically rare pathologies and diverse skin tones, scaling existing archives yields diminishing informational returns. 
Continued accumulation in high-density clusters (\textit{i.e.}, \gls{fst} II--III, \textit{common neoplasms}) may lead to minor diagnostic performance gains in arguably important cases, but contributes little to the model's ability to recognize rare or atypical presentations.

\subsection*{Evaluating dataset similarity and redundancy}\label{subsec:hidden_relationships}

The current dermatology open data landscape comprises 29 nominally distinct datasets. 
To quantify how distinct these datasets are in practice, we measured pairwise distances in the \textsc{SkinMap} embedding space as a proxy for mutual diversity. 
The resulting similarity density (Fig.~\ref{fig:Novelty_Dataset_Simi}d) is skewed toward low distances, implying that most public datasets differ only modestly and do not constitute clearly separable clinical populations. 
Consistent with this, we identify several high-overlap pairs (Fig.~\ref{fig:Novelty_Dataset_Simi}e) with near-identical statistical structure, including established relationships (\textit{e.g.}, \texttt{Fitzpatrick17k}--\texttt{SkinCAP} and \texttt{CAN2000}--\texttt{CAN5600}) and substantial similarities between annual challenge datasets and their parent institutional archives (\textit{e.g.}, \texttt{ISIC-challenge-2017}--\texttt{MSKCC}, \texttt{Derm1M}--\texttt{PubMed}, and \texttt{Derm1M}--\texttt{SkinCAP}). 
Although such overlap may be expected when data originate from shared institutions, it has direct implications for evaluation: Validation on a related dataset can masquerade as ``external'' testing while effectively reproducing internal validation, artificially inflating reported performance, and obscuring gaps in real-world generalization.

\looseness=-1
To guide future data selection, we propose a score for dataset uniqueness (Fig.~\ref{fig:Novelty_Dataset_Simi}f), computed as the average distance of a dataset to the global collection of dermatology. 
General-purpose datasets such as \texttt{Derm1M} are close to the statistical center of the map, indicating that they represent typical patient profiles currently available to \gls{ai}. 
Undoubtedly useful for teaching models the basics, adding more such data yields limited gains in model robustness. 
In contrast, datasets such as \texttt{PH2} and \texttt{MED-NODE} populate the outer regions of the map. 
\texttt{PH2} and \texttt{MED-NODE} contribute distributional variance through a tightly controlled acquisition protocol and lesion characteristics that are comparatively uncommon in larger archives.
These ``outliers'' are high-value assets: They contain the distinct, non-standard variations necessary to rigorously stress-test an \gls{ai}'s ability to handle diverse real-world populations.

\subsection*{Identifying structural gaps in the data landscape}\label{subsec:dark_corners}

\begin{figure}[htbp]
    \centering
    \includegraphics[width=\linewidth]{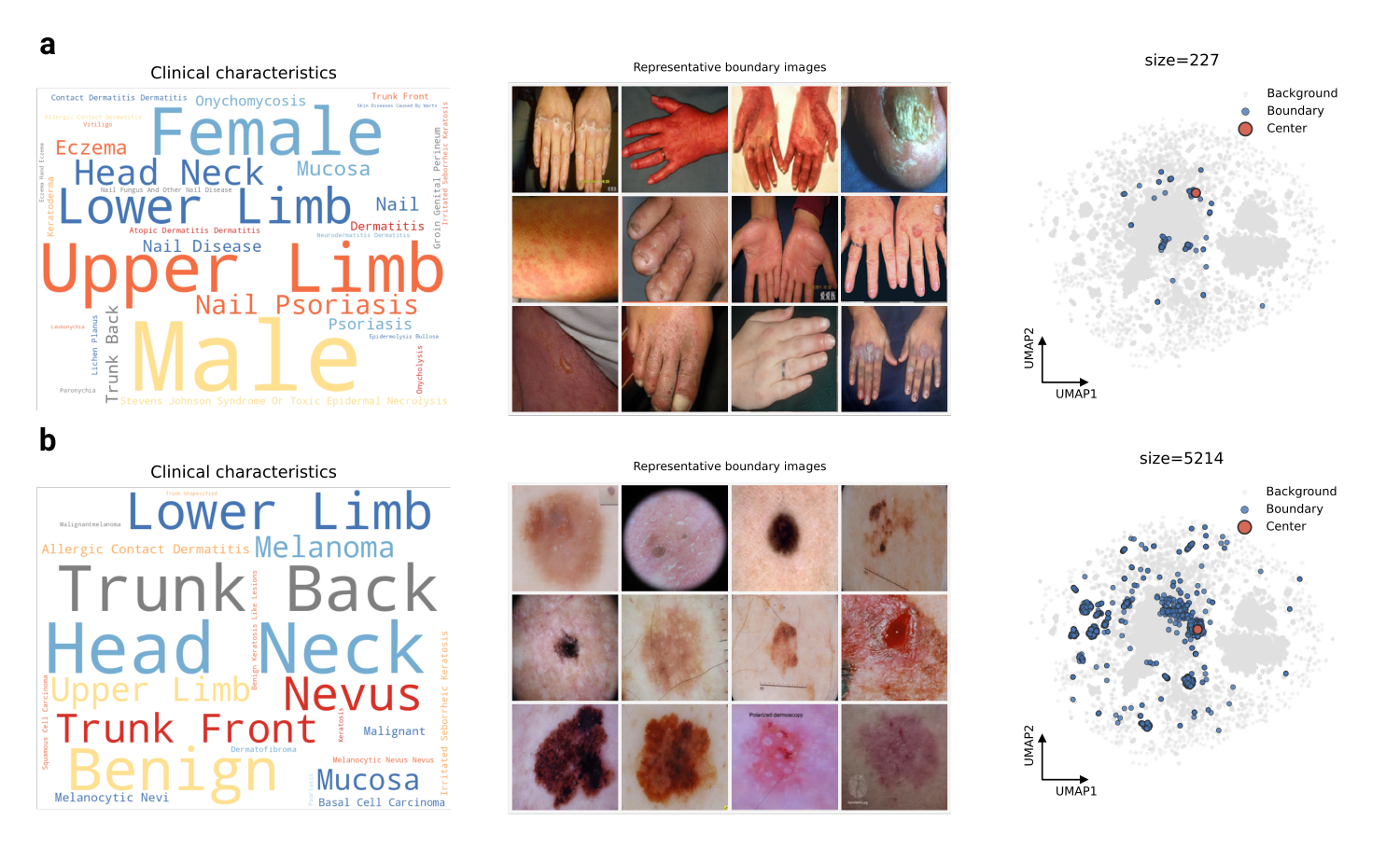}
    \caption{
    \textbf{Detection of topological holes in the embedding space.}
    \textbf{a}, Detailed analysis of the highest-persistence void. 
    The word cloud of metadata from boundary samples is dominated by \textit{nail}, \textit{nail disease}, and \textit{nail psoriasis}, confirming that this topological void represents a structural absence of nail pathologies. 
    \textbf{b}, Analysis of a second void bordered by diverse dermoscopy images of the \textit{lower limb}, \textit{head}, and \textit{neck}. 
    The existence of this gap indicates a structural lack of varying conditions and intermediate presentations within this modality in the current archive.
    }
    \label{fig:HoleDetectionResult}
\end{figure}

To assess the coverage of the global dermatological data collection, we analyze the density of different regions of the \textsc{SkinMap} embedding space. 
We identify regions of hypersaturation and sparsity, \textit{i.e.}, clusters with the highest and lowest 2.5\% density quantiles (Supplementary Fig.~\ref{fig:DensitySpace}), revealing that the atlas is polarized between massive redundancy and distinct structural voids.
The high-density regions represent statistically overrepresented visual clinical presentations (Supplementary Fig.~\ref{fig:DensitySpace}c). 
These regions are dominated by near-indistinguishable patches of skin or common moles, confirming that continued accumulation in these regions yields diminishing returns. 
Conversely, low-density regions often highlight data-quality issues such as images unrelated to dermatology, reconstruction artifacts, and heavily occluded images (Supplementary Fig.~\ref{fig:DensitySpace}d). 
Thus, embedding density can also serve as an automated quality-assurance mechanism, \textit{i.e.}, by pruning redundant clusters and filtering outliers.

Moreover, we applied spectral persistent homology~\citep{damrich2024persistent} to detect topological holes, \textit{i.e.}, void regions with almost no cases (Fig.~\ref{fig:HoleDetectionResult}). 
Unlike simple low-density regions, these holes represent ``unknown unknowns'': Valid clinical phenotypes that are structurally absent from the training data.
Crucially, this analysis independently confirms our metadata findings (Supplementary Information \ref{app:SparseMetadata}). 
We identify a topological hole specifically circumscribed by nail pathologies (Fig.~\ref{fig:HoleDetectionResult}a). 
While the boundary contains scattered examples of \textit{onychodystrophy} and \textit{nail psoriasis}, its existence indicates a systemic lack of variation in this anatomy. 
This confirms that the scarcity of nail conditions identified in our metadata audit is not merely a labelling issue, but a fundamental visual gap in the global atlas. 
By identifying these voids, \textsc{SkinMap} enables targeted data collection by specifying which phenotypes and demographic intersections are missing from current public archives. 
In the global dermatology data basis, high-priority gaps include nail disorders, pediatric dermatoses, especially in darker skin tones, and low-prevalence \gls{icd} blocks that appear only transiently in the public record.

\subsection*{Digital atlas}\label{subsec:clinical_usability}
\begin{figure}[htbp]
    \centering
    \includegraphics[width=\linewidth]{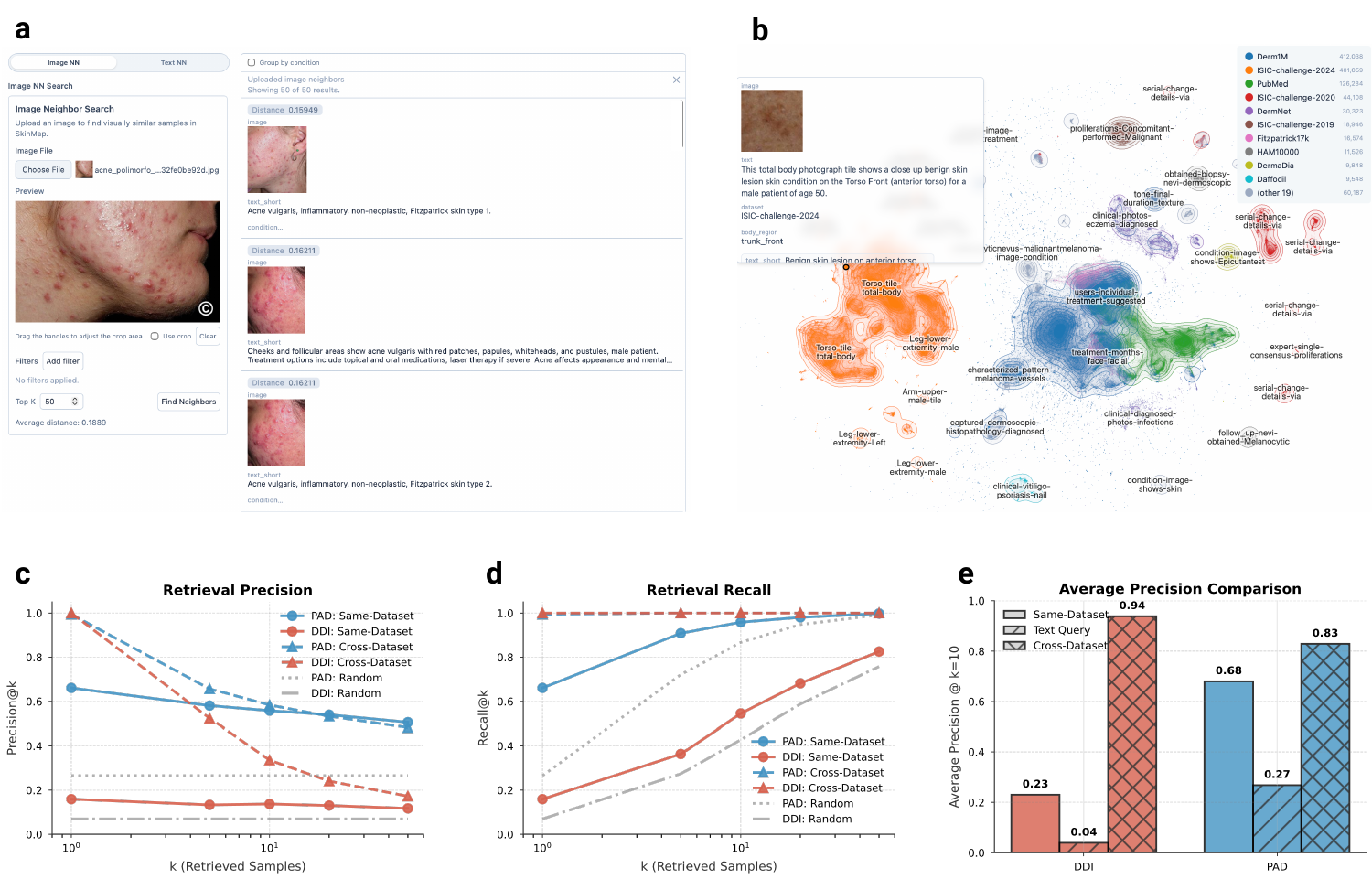}
    \caption{
    \textbf{The \textsc{SkinMap} digital atlas for clinical decision support and strategic data steering.}
    \textbf{a}, Interactive web-based interface allowing for real-time case uploads and retrieval of semantically similar cases with associated ground-truth diagnoses. 
    \textbf{b}, Visualization of the global latent space, projecting diverse datasets into a unified manifold where proximity corresponds to clinical similarity and allows practitioners to explore the atlas. 
    \textbf{c--d}, Benchmarking of retrieval precision (\textbf{c}) and recall (\textbf{d}) on hold-out datasets, comparing baseline performance within the same dataset against cross-dataset retrieval using the complete atlas. 
    \textbf{e}, Comparison of Average Precision (at $k\!=\!10$) demonstrating that while same-dataset retrieval provides a baseline for specialized cohorts like \texttt{DDI}, cross-dataset retrieval through the exhaustive \textsc{SkinMap} atlas provides a significant performance boost, increasing \texttt{DDI} precision from 0.23 to 0.94.
    }
    \label{fig:DigitalAtlas}
\end{figure}

To transform these findings from a static audit into an actionable resource, we developed a web-based digital atlas that enables the community to interact directly with the \textsc{SkinMap} embedding space.
This interactive platform serves two primary research purposes: Exploratory comparative analysis of dermatological presentations and strategic steering of data collection initiatives.
To support diagnostic workflows, the application allows practitioners to upload cases, which are projected in real time onto the shared latent manifold (Fig.~\ref{fig:DigitalAtlas}a--b). 
By leveraging the unified collection of 1.1~million images, the system retrieves and displays the most semantically similar cases along with their ground-truth diagnoses, also allowing for dynamic filtering of attributes such as body region or \gls{fst}. 
This capability effectively provides clinicians with a global reference standard, moving beyond the limitations of local institutional archives.

To validate the clinical utility of this digital atlas, we benchmarked retrieval performance using two hold-out datasets, \texttt{DDI} and \texttt{PAD-UFES-20}, measuring precision and recall across both same-dataset and cross-dataset (atlas-wide) settings (Fig.~\ref{fig:DigitalAtlas}c--e). 
Our results indicate that same-dataset retrieval remains a relatively effective baseline, even in the highly challenging, long-tailed disease distributions characteristic of the \texttt{DDI} dataset. 
However, the performance is fundamentally constrained by the limited sample volume of rare pathologies within single institutions.
By expanding the search to the complete \textsc{SkinMap} atlas, we observe a substantial performance boost due to the system's ability to find semantically similar ``neighbors'' across the exhaustive 1.1 million-image collection (Fig.~\ref{fig:DigitalAtlas}e). 
For the \texttt{DDI} dataset, average precision (at $k\!=\!10$) increases from 0.23 in the same-dataset setting to 0.94 using the full atlas. 
Similarly, \texttt{PAD-UFES-20} retrieval precision improves from 0.68 to 0.83. 
This capability provides clinicians with a global reference standard, effectively bridging local data gaps by leveraging the collective informational density of the entire dermatological landscape.

By further enabling practitioners to query the \textsc{SkinMap} embedding space and compute average distances to known collections and samples, the platform provides the necessary infrastructure for strategic steering of data acquisition. 
By establishing quantitative KPIs for dataset diversity, the atlas enables funding bodies and research consortia to optimize resource investment, prioritizing high-gain samples that fulfill target goals over guesswork.

\section*{Discussion}\label{sec:discussion}
For the past decade, the dermatology \gls{ai} community has operated under the assumption of data bias, yet the scale and topography of this bias, as well as the marginal novelty of new data contributions, have remained unquantified. 
By unifying over 1.1 million fragmented images into a queryable semantic atlas, this study provides a quantitative audit of the open data foundations of modern dermatological \gls{ai} systems. 
The findings expose a systemic divergence between the ``digital population'' used for model training and the global human population. 
Specifically, the representation of \gls{fst} V and VI is restricted to 5.8\% of the global dermatology collection, and pediatric data comprises only 3.0\%. 
Rather than a simple lack of samples, these statistics indicate a structural exclusion where models are developed on a demographic distribution that is heavily skewed toward Western, middle-aged, and lightly pigmented populations. 
This geographic concentration, primarily in the Global North, suggests that without targeted intervention, algorithmic performance will likely not generalize to the populations bearing a significant portion of the global disease burden.
Crucially, the temporal analysis challenges the prevailing assumption that dataset scale alone drives progress. 
The observed decoupling of data volume from informational novelty indicates that the strategy of indiscriminate data accumulation has reached a point of diminishing returns. 
Despite growth in dataset sizes, the novelty of new samples has plateaued. 
Similar patterns, in which output grows while disruptive novelty declines, have been reported across science and technology~\citep{funk2025disruption,park2023papers}, underscoring the need for explicit measures of marginal dataset contribution rather than relying solely on scale.
Together with the density and demographic analyses, these results indicate that recent growth has disproportionately accrued in already high-density regions, specifically \textit{benign neoplasms} on fair skin, whereas high-value, low-density regions, such as rare pathologies and diverse skin tones, remain stagnant. 
The persistence of these voids implies that continued aggregation of convenience samples will likely yield minimal improvements in model robustness and generalization.


Consequently, these findings advocate for a paradigm shift from volume-centric accumulation to strategic, precision data acquisition. 
\textsc{SkinMap} offers the infrastructure to operationalize this shift, serving as a mechanism to assess the marginal utility of potential data collections before resources are committed to annotation. 
By projecting new cohorts into a shared space, the community can distinguish between redundant contributions and high-priority samples that fill in topological voids. 
This metric-driven approach ensures that future infrastructure partnerships prioritize addressing structural gaps in the data landscape, thereby aligning the digital representation of dermatology with the diversity of real-world clinical practice.
We hope that \textsc{SkinMap} will be a catalyst for a collaborative effort to maintain a comprehensive collection of digital skin datasets, thus accelerating dermatology in both clinics and research institutions.
In practical terms, the atlas suggests a number of actionable acquisition directions: Expand geographically through infrastructure partnerships in underrepresented regions, build pediatric datasets under strengthened consent and privacy protections, increase coverage of underrepresented anatomical sites such as nails, and prioritize sustained acquisition of rare and non-neoplasm diagnostic categories to prevent ``orphan'' conditions from disappearing from the public record. 
These directions operationalize dataset innovation as measurable expansion of clinical coverage rather than continued densification of already-saturated clusters.

\section*{Methods}\label{sec:methods}

We built a global atlas of public dermatology imaging by (i) aggregating and harmonizing 29 datasets into a single corpus, (ii) converting standardized structured metadata into consistent image--text pairs, (iii) constructing the \textsc{SkinMap} embedding space via an ensemble of multi-modal and self-supervised encoders, (iv) imputing missing demographic and anatomical attributes using linear probes on frozen embeddings, and (v) auditing the resulting landscape with complementary metrics (\textit{i.e.}, temporal novelty, dataset similarity, density, and topology analyses).

\subsection*{Data collection and harmonization}
We conducted a comprehensive review and aggregation of the global public dermatology data landscape, identifying 29 publicly available datasets through a survey of biomedical repositories, challenge platforms, and published literature.
We included datasets that were publicly accessible for research use and contained dermatology images (\textit{i.e.}, clinical photography, dermoscopy, or \gls{tbp}) with structured labels (\textit{e.g.}, diagnosis) or textual description.
A complete list of included datasets is provided in the Supplementary Information~\ref{app:Datasets}.
All images were retrieved from the official hosting source, and we preserved the original dataset identifiers to maintain traceability throughout the pipeline.
To reduce duplicates, we performed identifier-based deduplication wherever reliable identifiers were available.
Specifically, we used dataset-provided image/lesion/case identifiers and, where present, patient identifiers to detect repeated samples within the same source and across re-releases of the same organizational archive.
For large multi-release archives (\textit{e.g.}, \texttt{ISIC} and related derivatives), we cross-referenced identifiers across the different subsets/challenges and removed duplicates, retaining a single canonical instance per identifier.
This effort yielded a collection of 1,135,082 unique images.
The aggregated corpus is heterogeneous, incorporating dermoscopic images, clinical photography, and \gls{tbp} from diverse sources such as institutional archives or educational web resources.
To use these disparate data sources for multi-modal training, we first manually harmonized the available structured metadata fields across datasets (\textit{i.e.}, diagnosis, body site, age, gender, and origin), resolving dataset-specific conventions and synonymy into a unified schema.
Body-site information and other free-text fields were manually mapped to a consistent controlled vocabulary (\textit{e.g.}, anatomical regions), and age was standardized to common formats and units.
After this harmonization step, we converted the structured metadata into natural-language descriptions to create consistent text-image pairs for multi-modal learning.
We utilized a template description followed by a \gls{llm} refinement step (Qwen2.5-7B-Instruct~\citep{qwen2,qwen2.5}) to generate concise, clinically coherent captions for all images, ensuring a consistent text-image pair format for subsequent contrastive learning (Supplementary Information~\ref{app:Prompt}).
The \gls{llm} step was constrained to stylistic normalization (\textit{e.g.}, brevity and consistent phrasing) while preserving the underlying structured content.
Additionally, dataset labels were manually assigned to \gls{icd}-10 codes whenever possible.
Note, the \gls{icd} codes are used solely in subsequent analyses, whereas the original labels are used for model training.

\subsection*{SkinMap embedding space}
To harmonize heterogeneous datasets into a single analyzable entity, we constructed the \textsc{SkinMap} embedding space using an ensemble learning approach that captures both high-level semantic concepts and low-level visual textures.
All images were embedded using a standardized preprocessing pipeline, fixed input resolution, and normalization consistent with each encoder.
The same pipeline was used for all analyses reported in this study.
We first trained multiple multi-modal models from scratch on curated image-text pairs using different objectives, such as CLIP~\citep{radford2021learning} and SigLIP~\citep{zhai2023sigmoid}, and on different partitions of the dataset.
Partitions were constructed to reduce dataset-specific shortcut learning (\textit{e.g.}, by mixing sources and holding out subsets during individual model training).
To leverage the full structural variance of the data without relying solely on possibly noisy metadata, we extended this with an ensemble of \gls{ssl} models, specifically DINO~\citep{caron_emerging_2021}, iBOT~\citep{zhou2022image}, and MAE~\citep{he2022masked}, pretrained on the imaging data alone.
For each encoder, we extracted a single global feature vector per image and $\ell_2$-normalized features before aggregation.
The final representation for each sample was generated by concatenating these diverse feature vectors and projecting them onto a shared, low-dimensional manifold via a learnable linear mapping, using a contrastive objective combined with the STRUCTURE regularization~\citep{groger2025limited} to preserve consistency with the pretrained space.
In practice, we used a fixed-dimensional representation for all analyses, \textit{e.g.}, 1024 dimensions.
Model and training hyperparameters, embedding dimensionalities for each backbone, and projector configuration are detailed in Supplementary Information~\ref{app:hyperparameters}.

\subsection*{Metadata imputation}
Given the sparsity of metadata in public collections, we used an imputation framework to estimate missing demographic labels.
We trained a suite of linear probes on the frozen \textsc{SkinMap} embedding space using the subset of samples with ground-truth annotations.
Separate probes were trained to predict each type of metadata to be recovered, namely origin, \gls{fst}, age, gender, and anatomical region.
This could be extended to other partially available information.
Each probe was implemented as a linear model trained on top of the frozen embedding using \texttt{scikit-learn}~\citep{scikit-learn}.
We used \texttt{LogisticRegression} for categorical targets and \texttt{Ridge} regression for continuous targets.
We validated the performance of the imputation engine on both an evaluation set and two external, distinct datasets (\texttt{DDI} and \texttt{PAD-UFES-20}) that were purposely excluded during pretraining.
Model performance was quantified using standard metrics for each attribute (\textit{e.g.}, macro-F1 for categorical labels and $R^2$ for continuous targets), and uncertainty was assessed via bootstrap confidence intervals.
Detailed performance comparison, including the evaluation of each model in the ensemble, is given in the Supplementary Table~\ref{tab:model_performance} and \ref{tab:model_performance_p2}.

\subsection*{Audit metrics}

\subsubsection*{Novelty over time}\label{app:yearly_novelty}
To assess diminishing returns in data acquisition, we defined a yearly novelty metric based on latent-space density. 
For each sample released in a given year, we computed the mean cosine distance to its $k$-nearest neighbors in the historical pool (all samples released in prior years).
To distinguish genuine informational gain from increasing dataset density, we constructed a null hypothesis using bootstrap resampling.
For each year, we generated a baseline distribution by randomly resampling points from the historical pool, matching the size of the new release, and computing their novelty scores. 
Observed novelty scores that converged toward this random baseline provided signal for saturation, indicating that new data releases were structurally redundant with existing archives.

Let $\{\mathbf{z}_i\}_{i=1}^N \subset \mathbb{R}^{d\times N}$ denote the samples' embeddings and $y_i \in \mathbb{Z}$ the associated release year.
We first $\ell_2$-normalized every embedding $\hat{\mathbf{z}}_i = \mathbf{z}_i/\|\mathbf{z}_i\|_2$, so cosine similarity between any two embeddings matched their inner product. 
For a fixed calendar year $t$, we defined the index sets
\begin{equation}
    \mathcal{I}_t = \{i : y_i = t\},
    \qquad
    \mathcal{P}_t = \{j : y_j < t\},
\end{equation}
corresponding to current-year samples $\mathcal{I}_t$ and the historical pool $\mathcal{P}_t$, respectively. 
For each sample $i \in \mathcal{I}_t$, we computed cosine similarities to $\mathcal{P}_t$, took the $k$ nearest neighbors, and converted them to distances,
\begin{equation}
    d_{i,\ell} = 1 - \hat{\mathbf{z}}_i^\top \hat{\mathbf{z}}_{\pi_{i,\ell}},
    \qquad 
    \ell = 1, \dots, k,
\end{equation}
where $\pi_{i,\ell}$ indexes the $\ell$-th neighbor of $i$ found via a cosine-similarity search.
Averaging the $k$ distances gave a per-sample novelty score
\begin{equation}
    \bar{d}_i = \frac{1}{k} \sum_{\ell=1}^k d_{i,\ell},
\end{equation}
and the raw yearly novelty was defined as the mean over all current-year samples,
\begin{equation}
    \nu_t = \frac{1}{|\mathcal{I}_t|} \sum_{i \in \mathcal{I}_t} \bar{d}_i.
\end{equation}

To disentangle true informational novelty from statistical artifacts of increasing pool sizes, we constructed a null hypothesis via bootstrap resampling.
For each year~$t$, we performed $B$ bootstrap iterations (with $B\!=\!200$).
In each iteration $b$, we randomly sampled (with replacement) $|\mathcal{I}_t|$ points from the historical pool $\mathcal{P}_t$ to form a pseudo-cohort $\mathcal{B}_t^{(b)}$.
We then computed their $k$-nearest neighbor distances within $\mathcal{P}_t$ (excluding self-matches), yielding a baseline novelty score
\begin{equation}
    \nu_t^{(b)} = \frac{1}{|\mathcal{I}_t|} \sum_{j \in \mathcal{B}_t^{(b)}} \bar{d}_j,
\end{equation}
where $\bar{d}_j$ is computed analogously to the observed case but with one additional neighbor to exclude the trivial self-match.
The empirical distribution $\{\nu_t^{(b)}\}_{b=1}^B$ characterizes the expected novelty under the null hypothesis that new samples are merely redrawn from previous years.
We summarize this via the mean
\begin{equation}
    \nu_t^\circ = \frac{1}{B} \sum_{b=1}^B \nu_t^{(b)}
\end{equation}
and construct confidence intervals using the $\alpha/2$ and $1-\alpha/2$ quantiles (here $\alpha=0.05$ for 95\% CI).
The ratio $\nu_t/\nu_t^\circ$ quantifies how much more novel the observed data is compared to random resampling, with values near 1 indicating saturation and values $>1$ indicating the population of previously uncovered regions.

\subsubsection*{Dataset similarity and uniqueness}\label{app:dataset_shift}
We quantified the similarity between distinct datasets using the \acrfull{fd} to assess shifts in means and covariances of their distributions.
For each dataset, we modeled its distribution in the \textsc{SkinMap} latent space as a multivariate Gaussian distribution with empirical mean and covariance. 
Specifically, pairwise \gls{fd} was computed across all 29 collections.
Pairs exhibiting low \gls{fd} (approaching zero) were identified as homologous, indicating that they occupy nearly identical latent manifolds and likely represent subsets or re-releases of the same data lineage. 
Conversely, we defined the uniqueness of a dataset as the average \gls{fd} of a collection relative to all others. 
This metric was used to identify high-value outliers that contribute significant distributional variance to the global atlas.

For each dataset $a$, we considered the normalized embeddings belonging to that dataset, $\{\hat{\mathbf{z}}_i : i \in \mathcal{D}_a\}$. 
We approximated the distribution with its empirical first and second moments:
\begin{equation}
    \boldsymbol{\mu}_a = \frac{1}{|\mathcal{D}_a|} \sum_{i \in \mathcal{D}_a} \hat{\mathbf{z}}
    _i,
    \qquad
    \Sigma_a = \frac{1}{|\mathcal{D}_a|-1} \sum_{i \in \mathcal{D}_a} (\hat{\mathbf{z}}_i -
    \boldsymbol{\mu}_a)(\hat{\mathbf{z}}_i - \boldsymbol{\mu}_a)^\top .
\end{equation}

Given any two datasets $a$ and $b$, we summarized their distributional discrepancy with the Fréchet distance between the corresponding Gaussians~\citep{dowson1982frechet},
\begin{equation}
    \operatorname{FD}(a,b) = 
    \|\boldsymbol{\mu}_a - \boldsymbol{\mu}_b\|_2^2 
    +\operatorname{Tr}\!\left( \Sigma_a + \Sigma_b - 2(\Sigma_a \Sigma_b)^{1/2} \right).
\end{equation}
This scalar captures both location shift (difference in means) and shape shift (covariance mismatch) between datasets under the assumed Gaussian approximation.

\subsubsection*{Density, redundancy, and hole detection}\label{app:HoleDetection}
To identify regions of redundancy and data quality failures within the landscape, we modeled the embedding space density using a Gaussian mixture model.
We defined operational thresholds at the 97.5th and 2.5th percentiles of log-density to partition the landscape into regions of hypersaturation and sparsity. 
High-density regions were analyzed to identify overrepresented visual concepts, while low-density regions were inspected to detect outliers, artifacts, and underrepresented clinical phenotypes.

To identify underrepresented regions (or \emph{holes}) in the embedding space, we employ a topological data analysis approach based on the spectral persistent homology method recently developed by~\citet{damrich2024persistent}. 
Traditional persistent homology methods fail in high-dimensional spaces with noise because ambient Euclidean distances become uninformative. \citet{damrich2024persistent} demonstrated that this limitation can be overcome by computing persistent homology on a filtered simplicial complex constructed from \emph{effective resistance distances} on the data manifold, which remain robust to high-dimensional noise by capturing the intrinsic geometry of the data.
We apply their method to the \textsc{SkinMap} embedding space to identify topological voids corresponding to underrepresented patient subgroups.

\bmhead{Effective resistance distance on $k$-nearest neighbor graphs}
Let $Z = \{\mathbf{z}_i\}_{i=1}^N \subset \mathbb{R}^d$ denote the set of embedding vectors in the $d$-dimensional latent space. 
We construct a $k$-nearest neighbor graph $G = (V, E)$ where vertices $V$ correspond to all data points and edges connect each point to its $k$ nearest neighbors.

The graph is symmetrized by including edge $(i, j)$ if $j$ is among the $k$ nearest neighbors of $i$ \emph{or} $i$ is among the $k$ nearest neighbors of $j$. 
Let $A \in \{0,1\}^{N \times N}$ denote the resulting adjacency matrix. We define the degree matrix $D = \text{diag}(d_1, \ldots, d_N)$ where $d_i = \sum_{j} A_{ij}$ and construct the graph Laplacian:
\begin{equation}
L = D - A.
\end{equation}

The \emph{effective resistance distance}~\citep{ghosh2008minimizing} between vertices $i$ and $j$ is defined through the pseudoinverse of the Laplacian. 
Let $L^\dagger$ denote the Moore--Penrose pseudoinverse of $L$, computed via eigendecomposition with regularization to handle the zero eigenvalue corresponding to the constant eigenvector. 
The effective resistance distance is given by:
\begin{equation}
d^\text{eff}_{ij} = (L^\dagger)_{ii} + (L^\dagger)_{jj} - 2(L^\dagger)_{ij}.
\label{eq:eff_resistance_naive}
\end{equation}

As noted by~\citet{hein2005graphs}, this naive formulation exhibits undesirable scaling properties in large graphs, where distances converge to approximately $1/d_i + 1/d_j$ independent of graph structure. Following~\citet{damrich2024persistent}, we apply the von Luxburg degree correction~\citep{hein2005graphs}, yielding the corrected effective resistance:
\begin{equation}
\tilde{d}^\text{eff}_{ij} = d^\text{eff}_{ij} - \frac{1}{d_i} - \frac{1}{d_j} + \frac{2A_{ij}}{d_i d_j} - \frac{A_{ii}}{d_i^2} - \frac{A_{jj}}{d_j^2}.
\label{eq:eff_resistance_corrected}
\end{equation}

The corrected distance $\tilde{d}^\text{eff}_{ij}$ captures the intrinsic manifold geometry by measuring the ``difficulty'' of random walks between vertices, remaining stable in the presence of high-dimensional noise~\citep{damrich2024persistent}. 
The resulting distance matrix $D^\text{eff} = [\tilde{d}^\text{eff}_{ij}]$ serves as input to the persistent homology computation.

\bmhead{Persistent homology and hole detection}
Given the distance matrix $D^\text{eff}$, we construct a Vietoris--Rips filtration~\citep{edelsbrunner2010computational}. 
A $k$-simplex $\sigma = \{v_{i_0}, \ldots, v_{i_k}\}$ is included in the complex at filtration parameter $\tau$ if and only if the maximum pairwise effective resistance distance among its vertices satisfies:
\begin{equation}
\max_{\ell, m \in \{0, \ldots, k\}} \tilde{d}^\text{eff}_{i_\ell i_m} \leq \tau.
\end{equation}

As $\tau$ increases from $0$ to $\infty$, simplices are added to the complex, forming a nested sequence of simplicial complexes (the filtration). 
We compute the persistent homology of this filtration using the Ripser algorithm~\citep{bauer2021ripser}, which efficiently extracts topological features across all scales.

A \emph{hole} in the embedding space corresponds to a 1-dimensional homology class (denoted $H_1$)~\citep{edelsbrunner2010computational}. 
Each such class persists over a filtration interval $[\tau_\text{birth}, \tau_\text{death}]$, where $\tau_\text{birth}$ is the scale at which a loop first appears and $\tau_\text{death}$ is the scale at which it is filled. 
The \emph{persistence} of the feature is defined as:
\begin{equation}
\text{persistence} = \tau_\text{death} - \tau_\text{birth}.
\end{equation}

Features with high persistence represent topologically significant structures, while low-persistence features typically correspond to noise~\citep{edelsbrunner2010computational}. 
We rank all detected $H_1$ features by their persistence values and select the top $k$ holes for further analysis, where $k$ is determined based on the dataset characteristics and the desired granularity of the gap analysis.

\bmhead{Cocycle extraction and hole characterization}
For each persistent $H_1$ feature, we extract its \emph{cocycle} representation~\citep{bauer2021ripser}: The set of edges (vertex pairs) that comprise the 1-cycle defining the hole. 
Let $\mathcal{C}_k = \{(i_1, j_1), (i_2, j_2), \ldots, (i_{|\mathcal{C}_k|}, j_{|\mathcal{C}_k|})\}$ denote the cocycle for the $k$-th hole. 
The unique vertices involved in this cocycle are:
\begin{equation}
\mathcal{V}_k = \{i : (i, j) \in \mathcal{C}_k \text{ or } (j, i) \in \mathcal{C}_k\}.
\end{equation}

These vertices represent the data points forming the boundary of the hole. 
To enable clinical interpretation and prioritization of detected gaps, we characterize each hole by:

\begin{itemize}
\item 
\textit{Hole center.}
The mean position in latent space of all vertices in the cocycle:
\begin{equation}
\mathbf{c}_k = \frac{1}{|\mathcal{V}_k|} \sum_{i \in \mathcal{V}_k} \mathbf{z}_i.
\label{eq:hole_center}
\end{equation}

\item
\textit{Hole size.}
The cardinality of the cocycle's vertex set: $s_k = |\mathcal{V}_k|$, representing the number of data points forming the hole.

\item
\textit{Hole volume.} 
We estimate the volume occupied by the hole in latent space by computing the empirical covariance of the cocycle vertices and modeling it as a $d$-dimensional hypersphere. Let:
\begin{equation}
r_k = \text{median}_{i \in \mathcal{V}_k} \|\mathbf{z}_i - \mathbf{c}_k\|
\end{equation}
denote the characteristic radius. 
The hole volume is approximated as:
\begin{equation}
V_k = \frac{\pi^{d/2}}{\Gamma(d/2 + 1)} r_k^d,
\label{eq:hole_volume}
\end{equation}
where $\Gamma$ is the gamma function.

\item
\textit{Boundary points.} 
To characterize the data surrounding each hole, we extract boundary samples by finding, for each vertex $i \in \mathcal{V}_k$, its $k_b$ nearest neighbors in the full dataset $Z$. 
We filter these neighbors by distance, retaining only those within:
\begin{equation}
\|\mathbf{z}_j - \mathbf{z}_i\| \leq \alpha \cdot \text{median}_{i' \in \mathcal{V}_k} d_{i', k_b},
\end{equation}
where $d_{i', k_b}$ is the distance to the $k_b$-th nearest neighbor of vertex $i'$ and $\alpha$ is a radius scale factor (here $\alpha = 1.5$). 
The union of these boundary points provides representative samples near the hole for qualitative inspection.
\end{itemize}


\backmatter

\bmhead{Acknowledgements}
We thank Martin A. Walter for their valuable suggestions, which helped improve the clarity of the manuscript.
This research was funded in part by the Swiss National Science Foundation (SNSF) under grant 20HW-1\_228541.
Figure~\ref{fig:Method} was created in \href{https://BioRender.com}{BioRender}.
All other figures were generated programmatically (Python/matplotlib) from the analyses described.

\section*{Declarations}
\begin{itemize}
\item \textbf{Funding:}
This work was supported in part by the Swiss National Science Foundation (SNSF; Grant 20HW-1\_228541).

\item \textbf{Competing interests:}
A.A.N. is a shareholder of Derma2Go AG and Derma.One GmbH. 
All other authors declare no competing interests.

\item \textbf{Ethics approval and consent to participate:}
Not applicable. 
This study used mainly publicly available datasets.
For the verification of the model predictions with domain experts (Fig.~\ref{fig:PerformanceComparison}), we solely collected answers with consent from the participants as categorical labels along with anonymized annotator identification. 
Thus, these annotations contain no personally identifiable information or
offensive content. 
In discussions with experts from the co-authors' institutions, it was concluded that this verification process does not require ethics approval because the study examines publicly available datasets and does not involve human subjects beyond categorical annotations.

\item \textbf{Consent for publication:}
Not applicable.

\item \textbf{Data availability:}
All datasets used in this study are publicly available from their original sources under their respective terms of use. 
Complete list of datasets and access information is provided in the Supplementary Information~\ref{app:Datasets}.
The datasets used in this study are publicly available and were de-identified by the original data curators prior to their release. 
The authors of this study had no access to personally identifiable information (PII) or patient keys.

\item \textbf{Materials availability:}
Resources associated with \textsc{SkinMap} (including derived artifacts such as the embedding space and metadata harmonization resources, where redistribution is permitted) will be released under an open-source license and made publicly available upon publication.

\item \textbf{Code availability:}
The source code for the \textsc{SkinMap} framework, including inference scripts, pretrained model weights, and the projection head parameters used to generate the embedding space, will be made available upon publication. 
The interactive digital atlas will be accessible upon publication.

\item \textbf{Author contributions:}
Conceptualization: F. Gröger; 
Methodology: F. Gröger, S. Lionetti; 
Software: F. Gröger; 
Validation: F. Gröger, S. Lionetti, P. Gottfrois; 
Formal analysis: F. Gröger, S. Lionetti; 
Investigation: F. Gröger, P. Gottfrois; 
Data curation: F. Gröger, P. Gottfrois; 
Writing -- original draft: F. Gröger; 
Writing -- review \& editing: All; 
Visualization: F. Gröger; 
Supervision: M. Groh, M. Pouly, A.A. Navarini; 
Funding acquisition: M. Pouly, A.A. Navarini.
\end{itemize}

\begin{appendices}
\clearpage
\section{Supplementary Figures}
\setcounter{figure}{0}
\renewcommand{\thefigure}{\arabic{figure}}
\renewcommand{\figurename}{Supplementary Fig.}
\begin{figure}[htbp]
    \centering
    \includegraphics[width=1.0\linewidth]{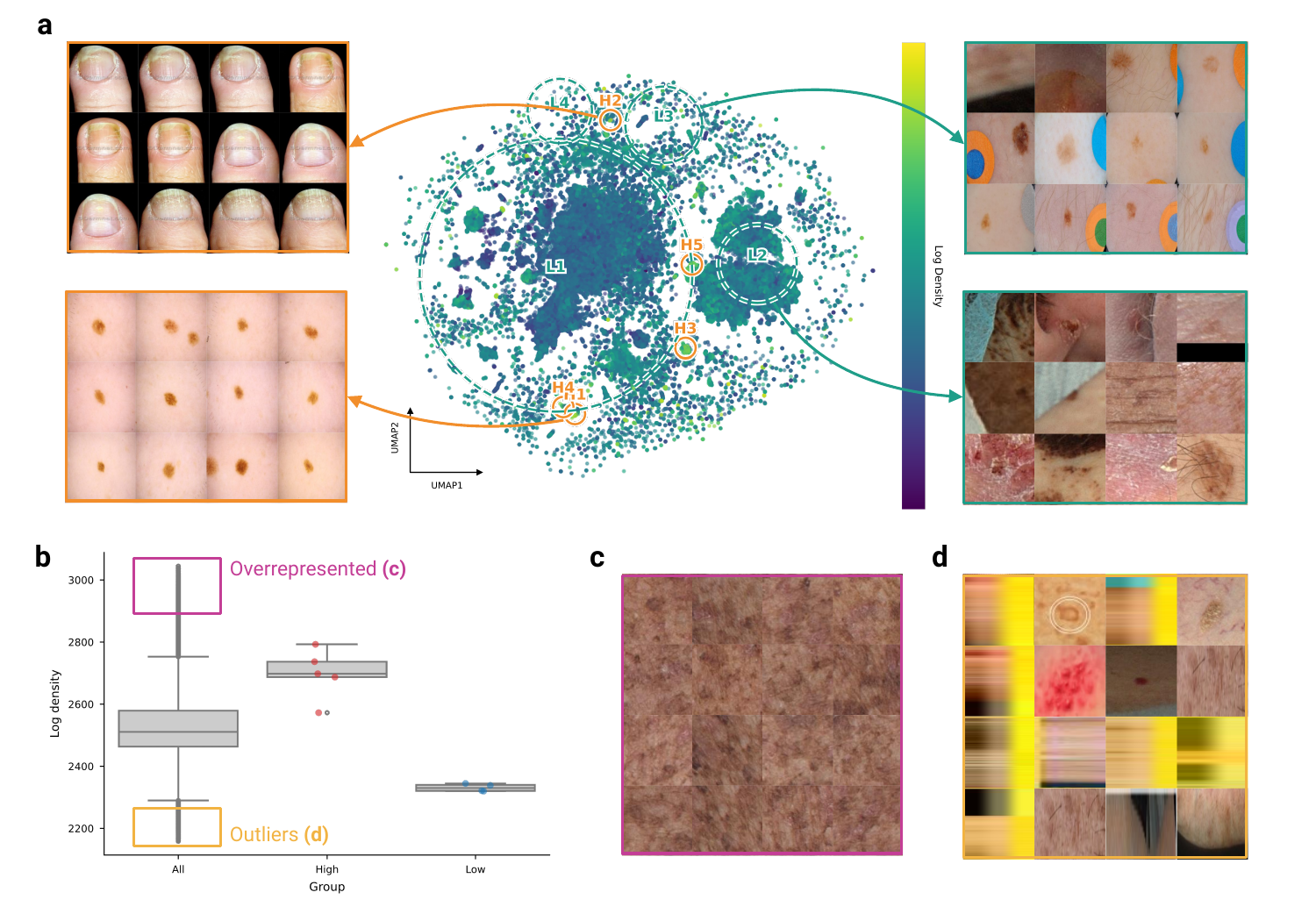}
    \caption{
    \textbf{Embedding space density as an automated quality assurance mechanism.} 
    \textbf{a}, UMAP projection of the \textsc{SkinMap} embedding space overlaid with a Gaussian mixture model density map. 
    Dashed circles indicate regions of hypersaturation (high density, $H1\!-\!H5$) and sparsity (low density, $L1\!-\!L4$). 
    \textbf{b}, Boxplot distribution of log-density scores distinguishing core clusters from peripheral outliers. 
    \textbf{c}, Representative samples from high-density peaks, revealing visual homogeneity and redundancy in common lesions. 
    \textbf{d}, Representative samples from low-density regions, effectively isolating data quality failures, including non-clinical images, reconstruction artifacts, and heavy occlusions.
    }
    \label{fig:DensitySpace}
\end{figure}

\begin{figure}[htbp]
    \centering
    \includegraphics[width=\linewidth]{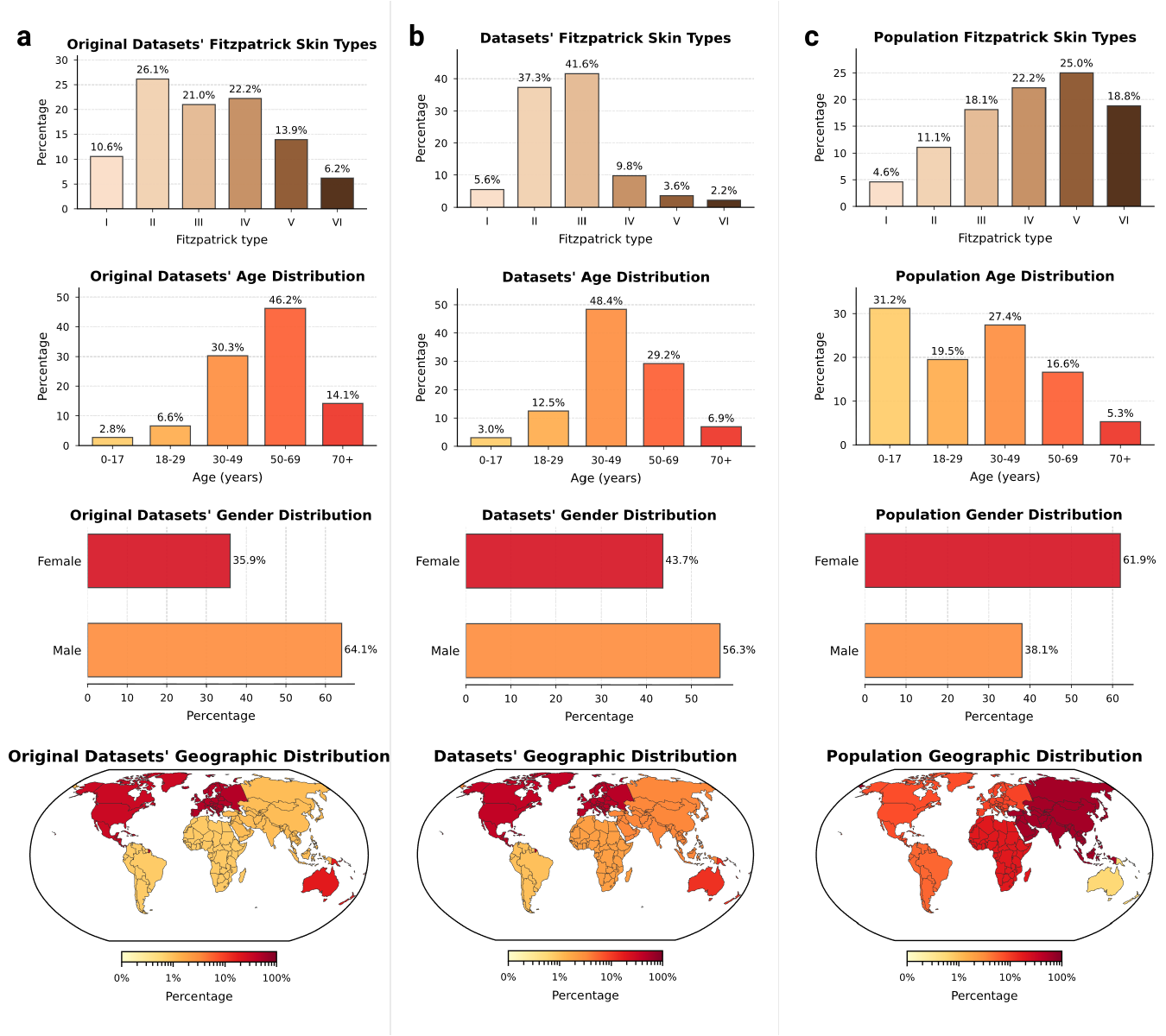}
    \caption{
    \textbf{Validation of imputed demographic distributions.} 
    Comparison of demographic distributions derived from original sparse metadata (left column, \textbf{a}), the fully imputed \textsc{SkinMap} attributes (center column, \textbf{b}), and global population baselines (right column, \textbf{c}). 
    The imputed distributions (center) effectively reconstruct the latent demographics of the dataset, revealing the true extent of exclusions (such as the scarcity of \gls{fst} V–VI and pediatric data) that are otherwise obscured by missing metadata in the original archives.
    }
    \label{fig:DataRealityCheckInclOriginal}
\end{figure}
\begin{figure}[htbp]
    \centering
    \includegraphics[width=0.95\linewidth]{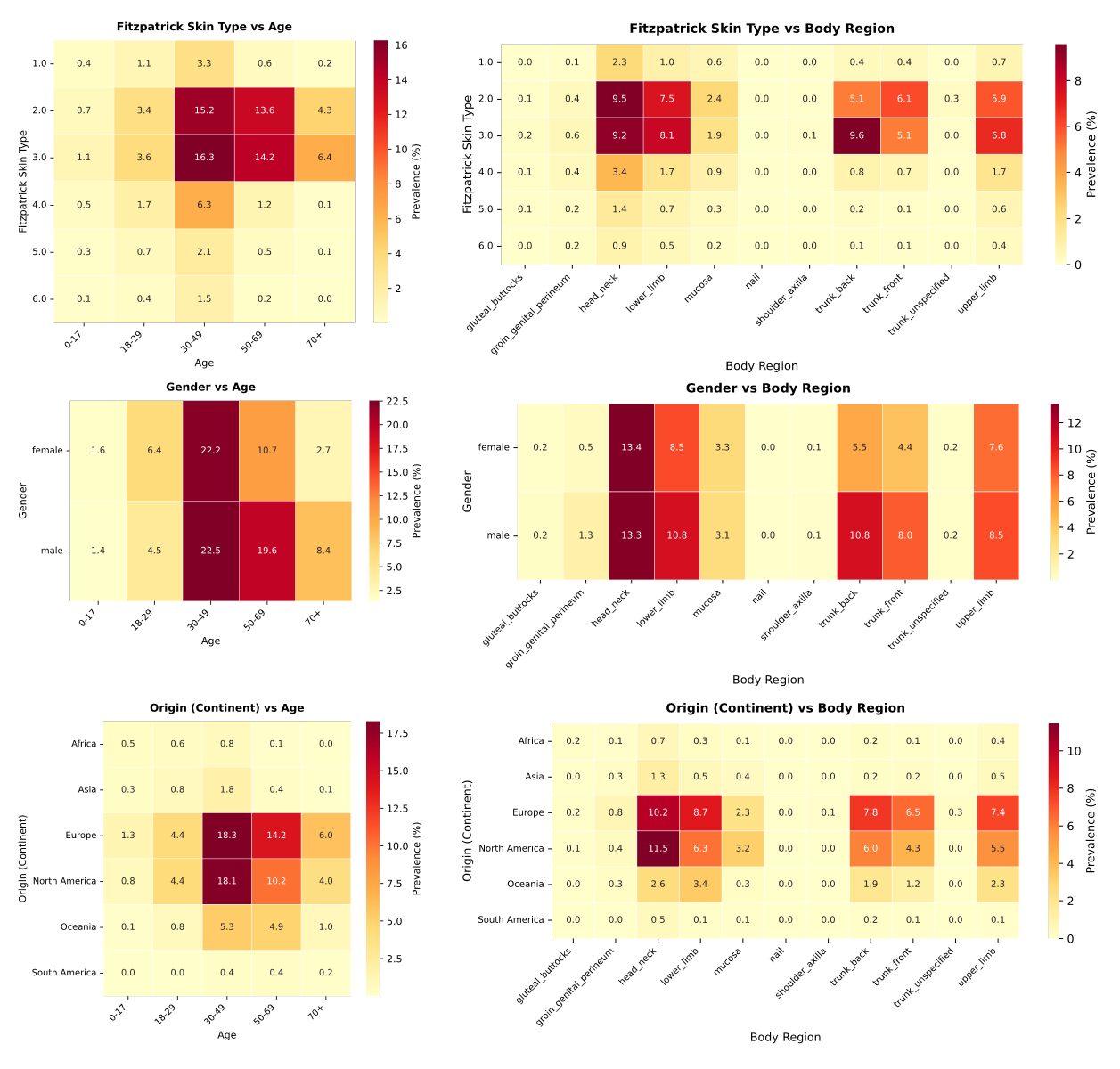}
    \caption{
    \textbf{Intersectional metadata audit reveals compounded underrepresentation in demographic subgroups.}
    Heatmaps displaying the prevalence of samples across intersecting demographic attributes (\textit{i.e.}, \gls{fst}, age, gender, geographic origin, and body region) in the global archive. 
    The color gradient represents the prevalence of samples, ranging from high saturation (red) to scarcity (yellow). 
    The analysis uncovers deep structural voids beyond single-axis statistics: 
    The interaction between skin tone and age highlights the critical lack of pediatric data for darker skin tones (\gls{fst} V-VI), where sample counts are negligible relative to adult \gls{fst} II populations. 
    Geographic origin versus skin tone confirms that, even when darker skin tones are present, they are often disconnected from their geographic centers of prevalence (\textit{e.g.}, the Global South) and instead originate from Western datasets. 
    Body region analysis reveals that non-standard anatomical locations on underrepresented demographics constitute ``dark corners'' of the latent space, potentially compromising diagnostic performance for complex clinical presentations. 
    These intersectional gaps indicate that global metrics of diversity mask profound local voids, where specific patient subgroups (\textit{e.g.}, elderly patients with \gls{fst} VI) remain effectively invisible to current \gls{ai} models.
    }
    \label{fig:PrevalenceForMeta}
\end{figure}
\begin{figure}[htbp]
    \centering
    \includegraphics[width=0.95\linewidth]{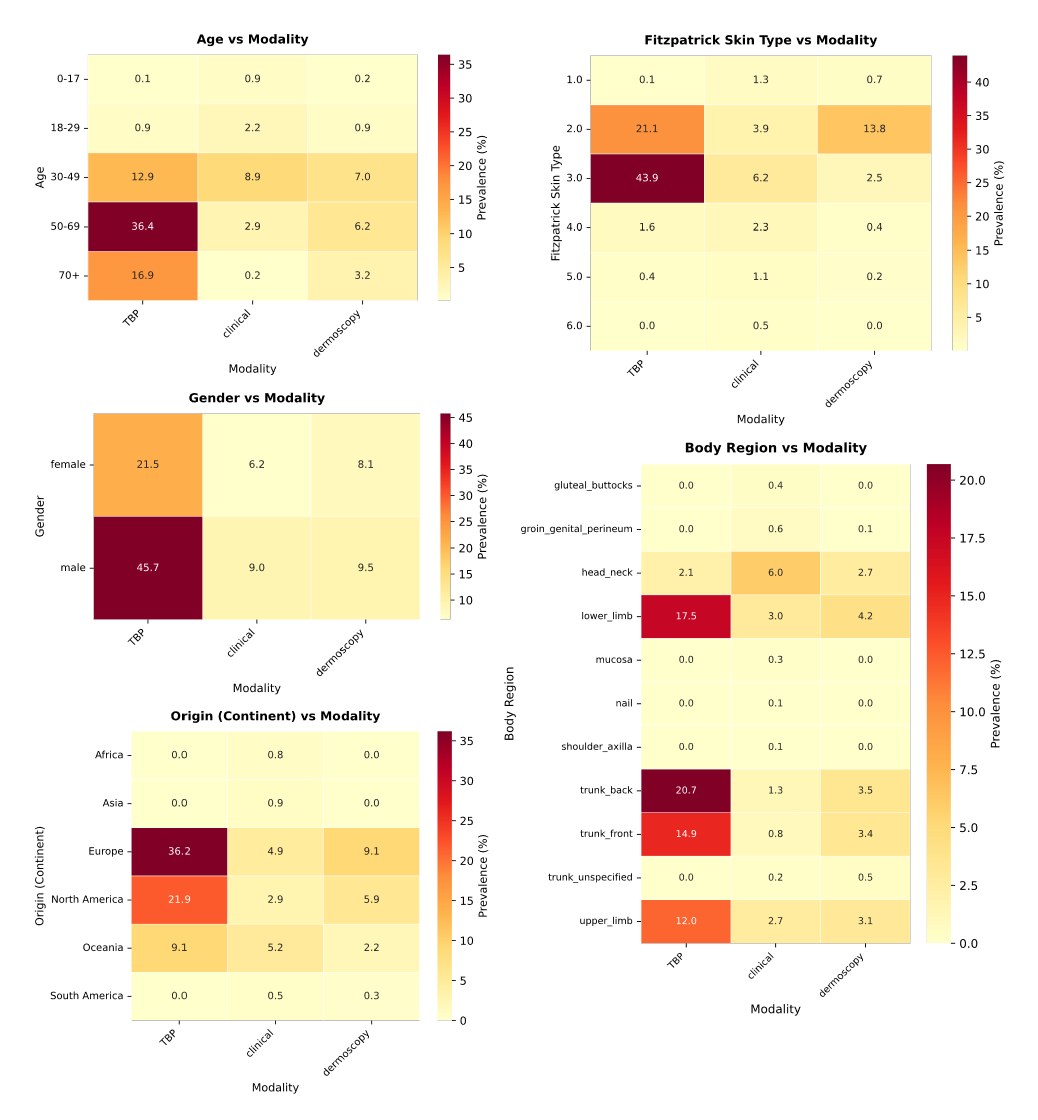}
    \caption{
    \textbf{Modality-specific distribution and structural bottlenecks. }
    Heatmaps displaying the prevalence of samples across intersections of imaging modality (\textit{i.e.}, \gls{tbp}, clinical, and dermoscopy) and demographic/clinical attributes. 
    The color gradient represents sample prevalence, from scarcity (yellow) to hypersaturation (red).
    }
    \label{fig:PrevalenceForMetaModality}
\end{figure}
\begin{figure}[htbp]
    \centering
    \includegraphics[width=1.0\linewidth]{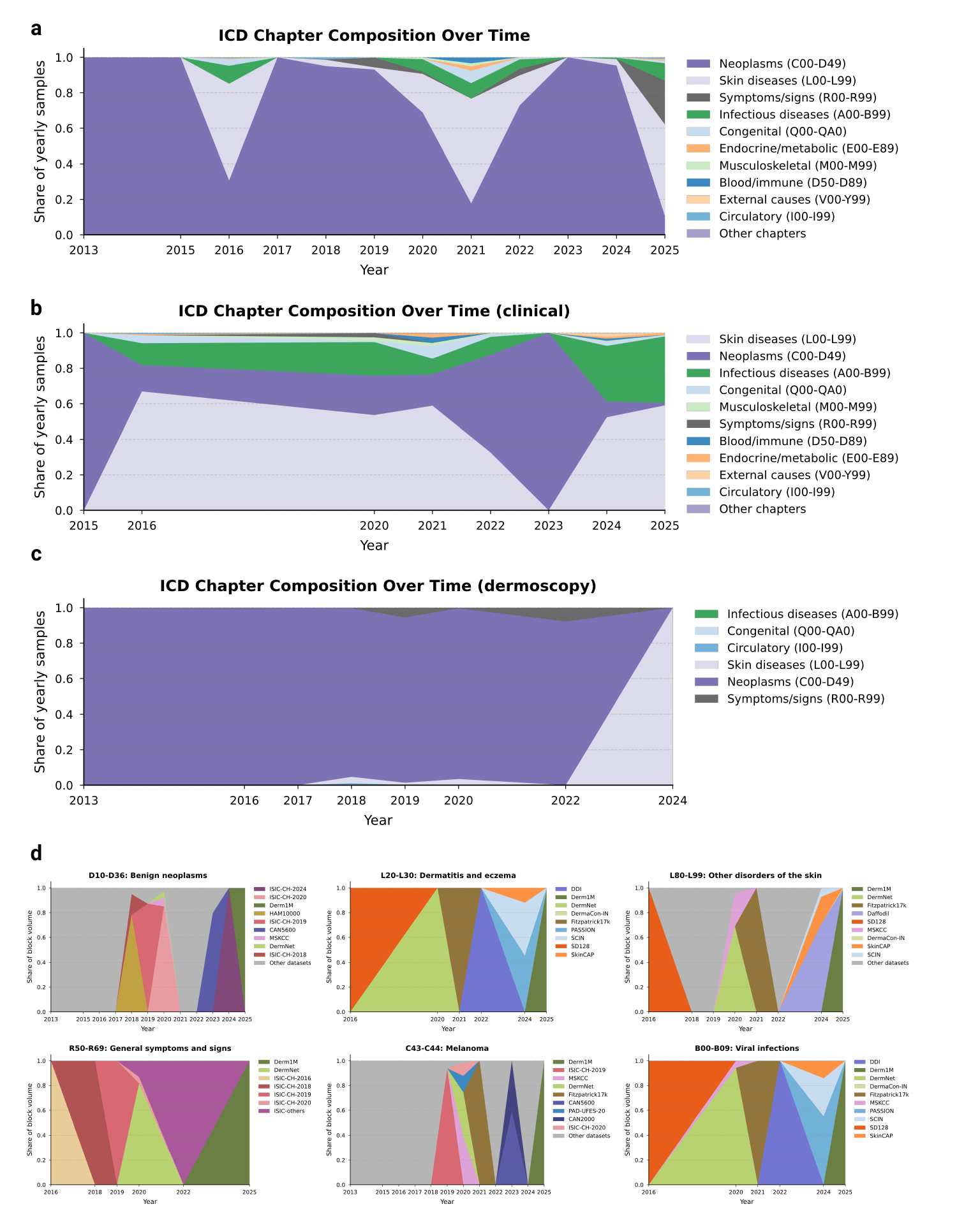}
    \caption{
    \textbf{Temporal evolution and provenance of \gls{icd}-10 disease categories.} 
    \textbf{a}, Stacked area chart displaying the proportional share of \gls{icd}-10 chapters in yearly data releases. 
    The landscape has become increasingly dominated by \textit{Neoplasms} (C00--D49) in recent years, correlating with the novelty saturation observed in the main analysis. 
    \textbf{b}, Granular provenance of specific diagnostic blocks (colors represent source datasets). 
    The visualization highlights that diversity is often driven by isolated, dataset-specific events (\textit{e.g.}, \texttt{ISIC challenges} that drive melanoma counts) rather than by broad, continuous acquisition.
    }
    \label{fig:ICDComposition}
\end{figure}
\begin{figure}[htbp]
    \centering
    \includegraphics[width=1.0\linewidth]{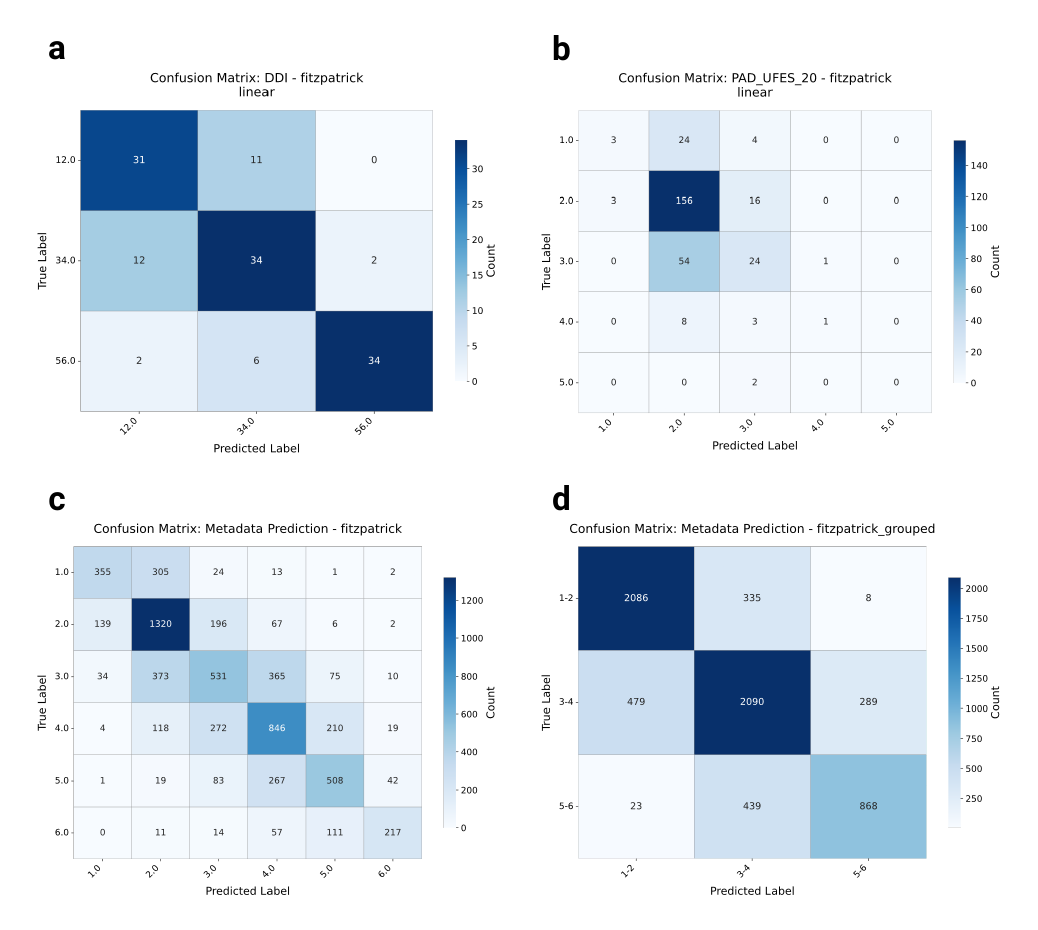}
    \caption{
    \textbf{Confusion matrices for Fitzpatrick skin type imputation.}
    Performance of the linear probe on \textbf{a}, the \texttt{DDI} external validation set; \textbf{b}, the \texttt{PAD-UFES-20} external validation set; and \textbf{c}, \textbf{d}, the internal hold-out test set for ungrouped (\textbf{c}) and grouped (\textbf{d}) skin tones. 
    Errors are concentrated around the diagonal (adjacent skin types), indicating that the model learns a coherent ordinal representation of skin tone rather than making random categorical errors.
    }
    \label{fig:ConfusionMatricesFST}
\end{figure}
\begin{figure}[htbp]
    \centering
    \includegraphics[width=1.0\linewidth]{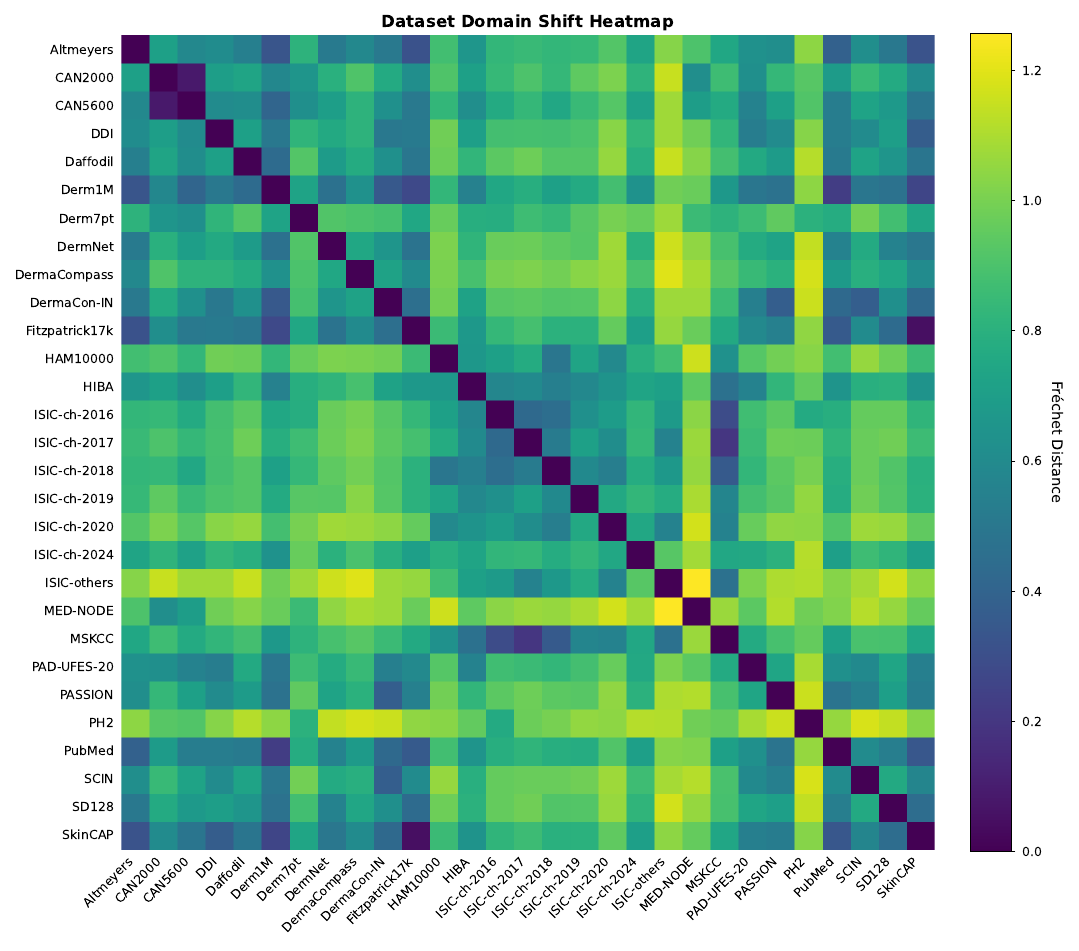}
    \caption{
    \textbf{Pairwise \acrfull{fd} heatmap quantifying inter-dataset domain shifts.}
    Matrix of \gls{fd} between all 29 aggregated datasets. 
    Dark blue indicates high similarity, statistically confirming subset relationships (\textit{e.g.}, \texttt{Fitzpatrick17k}--\texttt{SkinCAP}) and institutional clusters (\texttt{ISIC archives}). 
    Yellow/green bands identify high-uniqueness outliers that contribute significant distributional diversity to the global atlas.
    }
    \label{fig:DomainShiftHeatmap}
\end{figure}

\clearpage
\section{Supplementary Tables}
\setcounter{table}{0}
\renewcommand{\thetable}{\arabic{table}}
\renewcommand{\tablename}{Supplementary Table}

\begin{table}[ht]
\centering
\caption{
\textbf{Multi-modal backbone training hyperparameters.} 
All CLIP and SigLIP models were trained with these settings unless otherwise specified. 
}
\label{tab:clip_training_hyperparameters}
\begin{tabular}{ll}
\toprule
\textbf{Hyperparameter} & \textbf{Value} \\
\midrule
Optimizer & AdamW~\citep{loshchilov2017decoupled} \\
Base learning rate & $5 \times 10^{-6}$ \\
Weight decay & 0.01 \\
Batch size (per GPU) & 32 \\
Warmup ratio & 0.1 \\
LR schedule & Linear warmup + linear decay \\
Max gradient norm & 1.0 \\
Text token length & 77 \\
Mixed precision & FP16 \\
\bottomrule
\end{tabular}
\end{table}

\begin{table}[ht]
\centering
\caption{
\textbf{Model ensemble composition.} 
Each model contributes distinct feature representations to the final \textsc{SkinMap} space. 
}
\label{tab:model_ensemble}
\small
\begin{tabular}{lllll}
\toprule
\textbf{Architecture} & \textbf{Loss} & \textbf{Init} & \textbf{Epochs} & \textbf{Dataset} \\
\midrule
\multicolumn{5}{l}{\textit{Multi-modal models}} \\
 ViT-Large/14~\citep{dosovitskiy_image_2021} & CLIP~\citep{radford2021learning} & Random & 100 & 1/3 of dataset \\
 ViT-Large/14~\citep{dosovitskiy_image_2021} & CLIP~\citep{radford2021learning} & Random & 100 & 1/3 of dataset \\
 ViT-Large/14~\citep{dosovitskiy_image_2021} & CLIP~\citep{radford2021learning} & Random & 100 & 1/3 of dataset \\
 ViT-Large/14~\citep{dosovitskiy_image_2021} & SigLIP~\citep{zhai2023sigmoid} & Random & 100 & 1/3 of dataset \\
 ViT-Large/14~\citep{dosovitskiy_image_2021} & SigLIP~\citep{zhai2023sigmoid} & Random & 100 & 1/3 of dataset \\
 ViT-Large/14~\citep{dosovitskiy_image_2021} & SigLIP~\citep{zhai2023sigmoid} & Random & 100 & 1/3 of dataset \\
 ViT-Base/32~\citep{dosovitskiy_image_2021} & CLIP~\citep{radford2021learning} & Random & 100 & full dataset \\
 ViT-Base/32~\citep{dosovitskiy_image_2021} & SigLIP~\citep{zhai2023sigmoid} & Random & 100 & full dataset \\
\midrule
\multicolumn{5}{l}{\textit{Self-supervised models}} \\
 ViT-Base/16~\citep{dosovitskiy_image_2021} & DINO~\citep{caron_emerging_2021} & Random & 500 & full image dataset \\
 ViT-Base/16~\citep{dosovitskiy_image_2021} & iBOT~\citep{zhou2022image} & Random & 500 & full image dataset \\
 ViT-Base/16~\citep{dosovitskiy_image_2021} & MAE~\citep{he2022masked} & Random & 500 & full image dataset \\
\bottomrule
\end{tabular}
\end{table}

\begin{table}[ht]
\centering
\caption{
\textbf{Ensemble fusion and projector training hyperparameters. }
The projector maps concatenated multi-model embeddings to a unified 1024-dimensional representation.
}
\label{tab:projector_hyperparameters}
\begin{tabular}{ll}
\toprule
\textbf{Hyperparameter} & \textbf{Value} \\
\midrule
Projector architecture & Linear (single layer) \\
Output dimension & 1024 \\
Loss function & contrastive \\
STRUCTURE $\lambda$ & 10.0 \\
STRUCTURE hierarchy levels & 1 \\
STRUCTURE warmup steps & 500 \\
Training epochs & 20 \\
Learning rate & $5 \times 10^{-4}$ \\
Batch size & 512 \\
PCA initialization & true \\
\bottomrule
\end{tabular}
\end{table}

\begin{landscape}
\begin{table}[p] 
\centering
\caption{
\textbf{Comparative performance benchmarking of the \textsc{SkinMap} ensemble and foundation models.}
Detailed performance for attribute prediction across internal validation and external hold-out datasets (\texttt{DDI} and \texttt{PAD-UFES-20}). 
Metrics include macro-F1 scores for classification tasks and $R^2$ scores for regression-based attributes.
}
\label{tab:model_performance}
\begin{tabular}{lcccccc}
\toprule
\textbf{Model} & \textbf{DDI: disease} & \textbf{DDI: \gls{fst}} & \textbf{DDI: malignant} & \textbf{PAD: age} & \textbf{PAD: biopsed} & \textbf{PAD: bleed} \\
\midrule
Random & 0.01 & 0.38 & 0.56 & -0.53 & 0.49 & 0.49 \\
Ensemble & \textbf{0.09} & \textbf{0.76} & \textbf{0.60} & \textbf{0.40} & \textbf{0.88} & \textbf{0.75} \\
MONET & 0.05 & 0.72 & 0.54 & 0.37 & 0.84 & 0.72 \\
PanDerm & 0.04 & 0.74 & 0.45 & 0.39 & 0.86 & \textbf{0.75} \\
\midrule
ViT-L CLIP 1/3 & 0.05 & 0.67 & 0.55 & 0.28 & 0.80 & 0.66 \\
ViT-L CLIP 1/3 & 0.05 & 0.64 & 0.53 & 0.30 & 0.80 & 0.68 \\
ViT-L CLIP 1/3 & 0.04 & 0.68 & 0.56 & 0.24 & 0.82 & 0.64 \\
ViT-L SigLIP 1/3 & 0.05 & 0.66 & 0.55 & 0.30 & 0.78 & 0.63 \\
ViT-L SigLIP 1/3 & 0.04 & 0.59 & 0.54 & 0.18 & 0.74 & 0.58 \\
ViT-L SigLIP 1/3 & 0.04 & 0.57 & 0.58 & 0.21 & 0.77 & 0.63 \\
ViT-B CLIP & 0.05 & 0.73 & 0.54 & 0.39 & 0.87 & 0.68 \\
ViT-B SigLIP & 0.03 & 0.61 & 0.50 & 0.28 & 0.80 & 0.60 \\
ViT-B DINO & 0.04 & 0.65 & 0.50 & 0.38 & 0.83 & 0.73 \\
ViT-B iBOT & 0.05 & 0.73 & 0.58 & 0.39 & 0.80 & 0.63 \\
ViT-B MAE & 0.01 & 0.61 & 0.43 & 0.19 & 0.49 & 0.42 \\
\midrule
\midrule
\textbf{Model} & \textbf{PAD: changed} & \textbf{PAD: diagnostic n.} & \textbf{PAD: diameter} & \textbf{PAD: elevation} & \textbf{PAD: \gls{fst}} & \textbf{PAD: gender} \\
\midrule
Random & 0.48 & 0.15 & -0.58 & 0.50 & 0.20 & 0.50 \\
Ensemble & \textbf{0.55} & \textbf{0.67} & 0.24 & \textbf{0.77} & \textbf{0.29} & \textbf{0.77} \\
MONET & 0.47 & 0.48 & 0.19 & 0.74 & 0.16 & 0.69 \\
PanDerm & 0.47 & 0.47 & 0.23 & \textbf{0.77} & 0.15 & 0.75 \\
\midrule
ViT-L CLIP 1/3 & 0.47 & 0.43 & 0.14 & 0.66 & 0.17 & 0.64 \\
ViT-L CLIP 1/3 & 0.47 & 0.46 & 0.15 & 0.68 & 0.18 & 0.66 \\
ViT-L CLIP 1/3 & 0.47 & 0.44 & 0.11 & 0.66 & 0.18 & 0.62 \\
ViT-L SigLIP 1/3 & 0.47 & 0.45 & 0.12 & 0.69 & 0.18 & 0.67 \\
ViT-L SigLIP 1/3 & 0.47 & 0.36 & 0.00 & 0.61 & 0.23 & 0.54 \\
ViT-L SigLIP 1/3 & 0.47 & 0.41 & 0.04 & 0.66 & 0.20 & 0.58 \\
ViT-B CLIP & 0.47 & 0.59 & 0.20 & 0.75 & 0.16 & \textbf{0.77} \\
ViT-B SigLIP & 0.47 & 0.44 & 0.10 & 0.65 & 0.18 & 0.69 \\
ViT-B DINO & 0.47 & 0.55 & \textbf{0.28} & 0.74 & 0.16 & 0.71 \\
ViT-B iBOT & 0.47 & 0.45 & 0.25 & 0.71 & 0.17 & 0.71 \\
ViT-B MAE & 0.47 & 0.20 & 0.05 & 0.38 & 0.15 & 0.64 \\
\bottomrule
\end{tabular}
\end{table}
\end{landscape}

\begin{landscape}
\begin{table}[p] 
\centering
\caption{
\textbf{Comparative performance benchmarking of the \textsc{SkinMap} ensemble and foundation models (continued).}
Detailed performance for attribute prediction across internal validation and external hold-out datasets (\texttt{DDI} and \texttt{PAD-UFES-20}). 
Metrics include macro-F1 scores for classification tasks and $R^2$ scores for regression-based attributes.
}
\label{tab:model_performance_p2}
\begin{tabular}{lcccccc}
\toprule
\textbf{Model} & \textbf{PAD: grew} & \textbf{PAD: hurt} & \textbf{PAD: itch} & \textbf{PAD: region} & \textbf{SkinMap: age} & \textbf{SkinMap: body r.} \\
\midrule
Random & 0.49 & 0.50 & 0.52 & 0.08 & -0.61 & 0.09 \\
Ensemble & \textbf{0.73} & \textbf{0.75} & \textbf{0.76} & \textbf{0.42} & \textbf{0.74} & \textbf{0.66} \\
MONET & 0.71 & 0.62 & 0.74 & 0.19 & 0.42 & 0.50 \\
PanDerm & 0.72 & 0.61 & 0.75 & 0.15 & 0.51 & 0.49 \\
\midrule
ViT-L CLIP 1/3 & 0.64 & 0.55 & 0.71 & 0.16 & 0.58 & 0.57 \\
ViT-L CLIP 1/3 & 0.65 & 0.62 & 0.70 & 0.18 & 0.52 & 0.54 \\
ViT-L CLIP 1/3 & 0.69 & 0.56 & 0.70 & 0.15 & 0.57 & 0.56 \\
ViT-L SigLIP 1/3 & 0.67 & 0.58 & 0.70 & 0.16 & 0.52 & 0.53 \\
ViT-L SigLIP 1/3 & 0.63 & 0.50 & 0.65 & 0.10 & 0.65 & 0.59 \\
ViT-L SigLIP 1/3 & 0.67 & 0.59 & 0.67 & 0.12 & 0.52 & 0.54 \\
ViT-B CLIP & 0.72 & 0.60 & 0.74 & 0.22 & 0.59 & 0.58 \\
ViT-B SigLIP & 0.66 & 0.61 & 0.68 & 0.21 & 0.44 & 0.42 \\
ViT-B DINO & \textbf{0.73} & 0.66 & 0.72 & 0.40 & 0.45 & 0.49 \\
ViT-B iBOT & 0.71 & 0.57 & 0.71 & 0.33 & 0.45 & 0.49 \\
ViT-B MAE & 0.56 & 0.45 & 0.43 & 0.04 & 0.30 & 0.21 \\
\midrule
\midrule
\textbf{Model} & \textbf{SkinMap: \gls{fst}} & \textbf{SkinMap: \gls{fst} grp.} & \textbf{SkinMap: gender} & \textbf{SkinMap: origin} &  &  \\
\midrule
Random & 0.17 & 0.34 & 0.50 & 0.06 &  &  \\
Ensemble & \textbf{0.57} & \textbf{0.75} & \textbf{0.92} & \textbf{0.86} &  &  \\
MONET & 0.51 & 0.72 & 0.71 & 0.82 &  &  \\
PanDerm & 0.50 & 0.73 & 0.73 & 0.68 &  &  \\
\midrule
ViT-L CLIP 1/3 & 0.49 & 0.70 & 0.83 & 0.79 &  &  \\
ViT-L CLIP 1/3 & 0.50 & 0.70 & 0.78 & 0.81 &  &  \\
ViT-L CLIP 1/3 & 0.50 & 0.70 & 0.82 & 0.80 &  &  \\
ViT-L SigLIP 1/3 & 0.50 & 0.70 & 0.78 & 0.81 &  &  \\
ViT-L SigLIP 1/3 & 0.49 & 0.67 & 0.90 & 0.79 &  &  \\
ViT-L SigLIP 1/3 & 0.49 & 0.68 & 0.81 & 0.79 &  &  \\
ViT-B CLIP & 0.54 & 0.73 & 0.83 & 0.84 &  &  \\
ViT-B SigLIP & 0.45 & 0.65 & 0.74 & 0.70 &  &  \\
ViT-B DINO & 0.51 & 0.71 & 0.71 & 0.76 &  &  \\
ViT-B iBOT & 0.52 & 0.72 & 0.71 & 0.78 &  &  \\
ViT-B MAE & 0.34 & 0.62 & 0.63 & 0.38 &  &  \\
\bottomrule
\end{tabular}
\end{table}
\end{landscape}

\begin{landscape}
\begin{table}[p] 
\centering
\caption{
\textbf{Statistical reliability and confidence intervals (CI) for baseline model performance.}
Confidence intervals were generated using $B\!=\!1,000$ bootstrap samples. 
Data are presented as point estimate (95\% CI lower–upper). 
Performance is measured in terms of F1 for categorical labels and $R^2$ for continuous variables.
}
\label{tab:model_performance_ci}
\begin{tabular}{lcccccc}
\toprule
\textbf{Model} & \textbf{DDI: disease} & \textbf{DDI: \gls{fst}} & \textbf{DDI: malignant} & \textbf{PAD: age} & \textbf{PAD: biopsed} & \textbf{PAD: bleed} \\
\midrule
Ensemble & 0.09 (0.06--0.12) & 0.76 (0.69--0.83) & 0.60 (0.50--0.70) & 0.40 (0.28--0.45) & 0.88 (0.85--0.91) & 0.75 (0.71--0.80) \\
MONET & 0.05 (0.03--0.08) & 0.72 (0.65--0.80) & 0.54 (0.45--0.63) & 0.37 (0.27--0.44) & 0.84 (0.80--0.87) & 0.72 (0.67--0.77) \\
PanDerm & 0.04 (0.02--0.06) & 0.74 (0.67--0.81) & 0.45 (0.40--0.52) & 0.39 (0.29--0.46) & 0.86 (0.83--0.89) & 0.75 (0.70--0.80) \\
\midrule
\midrule
\textbf{Model} & \textbf{PAD: changed} & \textbf{PAD: diagnostic n.} & \textbf{PAD: diameter} & \textbf{PAD: elevation} & \textbf{PAD: \gls{fst}} & \textbf{PAD: gender} \\
\midrule
Ensemble & 0.55 (0.49--0.63) & 0.67 (0.60--0.72) & 0.24 (0.08--0.35) & 0.77 (0.73--0.80) & 0.29 (0.23--0.39) & 0.77 (0.72--0.82) \\
MONET & 0.47 (0.46--0.48) & 0.48 (0.44--0.51) & 0.19 (0.09--0.28) & 0.74 (0.70--0.78) & 0.16 (0.14--0.21) & 0.69 (0.64--0.75) \\
PanDerm & 0.47 (0.46--0.48) & 0.47 (0.44--0.51) & 0.23 (0.09--0.35) & 0.77 (0.73--0.82) & 0.15 (0.14--0.19) & 0.75 (0.70--0.80) \\
\midrule
\midrule
\textbf{Model} & \textbf{PAD: grew} & \textbf{PAD: hurt} & \textbf{PAD: itch} & \textbf{PAD: region} & \textbf{SkinMap: age} & \textbf{SkinMap: body r.} \\
\midrule
Ensemble & 0.73 (0.66--0.74) & 0.75 (0.70--0.81) & 0.76 (0.72--0.78) & 0.42 (0.35--0.49) & 0.74 (0.74--0.75) & 0.66 (0.65--0.67) \\
MONET & 0.71 (0.66--0.76) & 0.62 (0.55--0.68) & 0.74 (0.70--0.78) & 0.19 (0.16--0.22) & 0.42 (0.42--0.43) & 0.50 (0.49--0.51) \\
PanDerm & 0.72 (0.67--0.77) & 0.61 (0.55--0.67) & 0.75 (0.71--0.79) & 0.15 (0.13--0.17) & 0.51 (0.50--0.51) & 0.49 (0.47--0.50) \\
\midrule
\midrule
\textbf{Model} & \textbf{SkinMap: \gls{fst}} & \textbf{SkinMap: \gls{fst} grp.} & \textbf{SkinMap: gender} & \textbf{SkinMap: origin} &  &  \\
\midrule
Ensemble & 0.57 (0.55--0.58) & 0.75 (0.74--0.76) & 0.92 (0.92--0.93) & 0.86 (0.86--0.86) &  &  \\
MONET & 0.51 (0.49--0.52) & 0.72 (0.71--0.73) & 0.71 (0.70--0.71) & 0.82 (0.80--0.84) &  &  \\
PanDerm & 0.50 (0.48--0.51) & 0.73 (0.72--0.74) & 0.73 (0.72--0.73) & 0.68 (0.66--0.70) &  &  \\
\bottomrule
\end{tabular}
\end{table}
\end{landscape}

\begin{landscape}
\begin{table}[p] 
\centering
\small
\setlength{\tabcolsep}{5pt}
\renewcommand{\arraystretch}{1.15}
\caption{
    \textbf{Orphan ICD codes and transient acquisition patterns.}
    List of diagnostic categories that exhibit an episodic trajectory, appearing in specific collection windows (\textit{e.g.}, 2016 or 2020) but absent from subsequent acquisition efforts. 
    The collection years column highlights the temporal discontinuity of these samples, indicating a lack of sustained surveillance for pathologies outside the dominant clusters of benign and malignant neoplasms.
}
\label{tab:OrphanICD}
\begin{tabularx}{\linewidth}{@{} l L L S[table-format=4] c @{}}
\toprule
\textbf{ICD code} & \textbf{ICD description} & \textbf{ICD chapter} & {\textbf{Samples (n)}} & \textbf{Collection years} \\
\midrule
B09   & Unspecified viral infection characterized by skin and mucous membrane lesions
      & Certain infectious and parasitic diseases (A00--B99) & 1308 & 2020--2020 \\
A64   & Unspecified sexually transmitted disease
      & Certain infectious and parasitic diseases (A00--B99) & 436  & 2020--2020 \\
L23.7 & Allergic contact dermatitis due to plants
      & Diseases of the skin (L00--L99) & 325  & 2020--2020 \\
A38   & Scarlet fever
      & Certain infectious and parasitic diseases (A00--B99) & 35   & 2020--2020 \\
C7B.04& Secondary carcinoid tumors of peritoneum
      & Neoplasms (C00--D49) & 34   & 2020--2020 \\
B30.3 & Acute epidemic hemorrhagic conjunctivitis (enteroviral)
      & Certain infectious and parasitic diseases (A00--B99) & 32   & 2020--2020 \\
L66.2 & Folliculitis decalvans
      & Diseases of the skin (L00--L99) & 15   & 2020--2020 \\
L60.2 & Onychogryphosis
      & Diseases of the skin (L00--L99) & 11   & 2020--2020 \\
Q89.7 & Multiple congenital malformations, not elsewhere classified
      & Congenital malformations, deformations, chromosomal abnormalities, and genetic disorders (Q00--QA0) & 8 & 2020--2020 \\
L66.0 & Pseudopelade
      & Diseases of the skin (L00--L99) & 5    & 2020--2020 \\
E88.3 & Tumor lysis syndrome
      & Endocrine, nutritional and metabolic diseases (E00--E89) & 4    & 2020--2020 \\
L68.3 & Polytrichia
      & Diseases of the skin (L00--L99) & 2    & 2020--2020 \\
L81.2 & Freckles
      & Diseases of the skin (L00--L99) & 1    & 2020--2020 \\
I78.9 & Disease of capillaries, unspecified
      & Diseases of the circulatory system (I00--I99) & 177  & 2018--2018 \\
L64   & Androgenic alopecia
      & Diseases of the skin (L00--L99) & 41   & 2016--2016 \\
L85.9 & Epidermal thickening, unspecified
      & Diseases of the skin (L00--L99) & 40   & 2016--2016 \\
\bottomrule
\end{tabularx}
\end{table}
\end{landscape}

\clearpage
\section{Supplementary Information}
\subsection{Extended analysis}

\subsubsection{Sparse metadata against imputed}\label{app:SparseMetadata}

To validate that our audit of the global landscape reflects the true underlying demographics of the data rather than artifacts of the imputation process, we compared the distributional characteristics of the originally annotated subset with those of the fully imputed archive. 
Public medical datasets frequently suffer from missing metadata. 
In our aggregation, a substantial fraction of the 1.1 million images lacked definitive labels for key attributes such as \gls{fst}, age, and gender.

As illustrated in Supplementary Fig.~\ref{fig:DataRealityCheckInclOriginal}, we observe a strong concordance between the distributions derived from the sparse, ground-truth metadata (left column) and those generated by our imputation ensemble across the entire dataset (center column). 
For instance, the dominance of \gls{fst} II--III is consistent between the original subset and the full imputed atlas, as is the marginalization of \gls{fst} V--VI. 
Similarly, the pediatric population (ages 0--17) remains low in both the partial ($2.8\%$) and full ($3.0\%$) views, confirming that the unlabelled majority of the data does not harbor a hidden reservoir of diversity.

This structural consistency implies that the ``missingness'' of metadata in public archives is not driving the observed disparities. 
Rather, the bias is intrinsic to the image acquisition process itself. 
The unlabelled data essentially reflect the same exclusionary patterns as the labelled data: A heavy skew toward Western, middle-aged, and lightly pigmented populations. 
Consequently, the stark divergence from global population statistics (right column) reflects a genuine, systemic failure of the current dermatological data landscape to capture the diversity of the human population.

\subsubsection{Missing data from a metadata perspective}

While the univariate analysis presented in the main text exposes severe imbalances, such as the restriction of \gls{fst} V--VI to only 5.8\% of the archive and pediatric patients to 3.0\%, these aggregate statistics potentially mask even deeper intersectional voids. 
To investigate the availability of data for specific patient subgroups, we analyzed the joint distributions of imputed metadata attributes (Supplementary Fig.~\ref{fig:PrevalenceForMeta}).
This intersectional audit reveals that the global archive is not merely skewed, but highly concentrated around a specific ``standard'' patient profile. 
As illustrated in the \gls{fst} vs. age and gender vs. age heatmaps, the vast majority of the 1.1 million samples are clustered in a narrow demographic window: Patients aged 30--70 with \gls{fst}~II--III. 
This region of hypersaturation suggests that model performance reported in current literature is heavily weighted toward this specific demographic intersection.
Conversely, the analysis highlights ``dark corners'' of the latent space where data support collapses entirely. 
Most critically, the intersection of age and skin tone reveals a compounded exclusion of pediatric populations with darker skin tones. 
While pediatric data is scarce overall, the bin representing pediatric patients (0--17 years) with \gls{fst} V--VI is effectively empty. 
This implies that current foundation models possess virtually no ground-truth visual signals for common childhood dermatoses on darker skin, rendering them potentially unsafe for this vulnerable subgroup.
Furthermore, the origin (continent) vs. age analysis underscores the geographical homogeneity of the data. 
Even when diversity in age or body region is present, it is almost exclusively derived from datasets originating from the Global North (specifically Europe and North America). 
The distinct lack of samples from the Global South (Africa and South America) across all body regions suggests that environmental and genetic variations specific to these populations are absent from the training dataset.
Ultimately, these intersectional voids define the strict operational limits of the current Digital Atlas. 
Outside the saturated center of Western, middle-aged, fair-skinned patients, the reliability of diagnostic \gls{ai} remains unsupported by the underlying data.
The technical modality of data acquisition further compounds these structural exclusions (Supplementary Fig.~\ref{fig:PrevalenceForMetaModality}). 
While the recent expansion of the atlas has been driven by the introduction of \gls{tbp}, the intersectional analysis reveals that this modality is almost exclusively restricted to adult, fair-skinned populations (\gls{fst} II--III) within Europe and North America. 
Specifically, \gls{tbp} shows negligible coverage of pediatric patients and individuals with darker skin tones (\gls{fst} V--VI), whereas it is heavily concentrated on the trunk and lower limbs. 

\subsubsection{ICD composition over dataset}

To investigate the drivers behind the diminishing novelty observed in recent years, we analyzed the temporal evolution of disease categories and their specific data sources. 
Supplementary Fig.~\ref{fig:ICDComposition}a--c illustrates a heavy and persistent skew toward \textit{neoplasms} (C00--D49), particularly in the most recent acquisition phase (2022--2025), where this category dominates the vast majority of yearly samples. 
While earlier years display transient spikes in diversity, likely driven by specific releases containing \textit{symptoms/signs} (R00--R69) or \textit{skin diseases} (L00--L99), the overall trend demonstrates a consolidation around well-represented pathologies. 
Supplementary Fig.~\ref{fig:ICDComposition}d further granularizes this by mapping the provenance of specific diagnostic blocks, revealing that the availability of these categories is often contingent on isolated dataset release events rather than continuous collection. 
We observe that high-volume categories such as \textit{benign neoplasms} (D10--D36) and \textit{melanoma} (C43--C44) are heavily dictated by the sequential release of \texttt{ISIC} challenges and the \texttt{Derm1M} archive. 
This dependency highlights a fragility in the global data landscape, where the representation of entire disease categories rests on single sources rather than a robust, diverse network of data collectors.

\subsubsection{Orphan ICD codes}
Beyond the broad saturation of common neoplasms, our analysis uncovers a phenomenon of ``orphan'' diagnostic categories, \textit{i.e.}, pathologies that exhibit a transient presence in the global archive before vanishing from subsequent data collection efforts. 
As detailed in Supplementary Table~\ref{tab:OrphanICD}, this pattern effectively represents lost knowledge, where specific diseases appear during a single release event (often linked to a specific dataset such as \texttt{PAD-UFES-20} or disparate \texttt{ISIC challenges}) but are not sustained in longitudinal acquisition.
This discontinuity is most visible in infectious and inflammatory categories, which stand in stark contrast to the persistent accumulation of melanoma and nevi. 
For instance, \textit{unspecified viral infections} (B09) and \textit{sexually transmitted diseases} (A64) appeared as cohorts in 2020 ($n\!=\!1,308$ and $n\!=\!436$, respectively) but have seen no replenishment in the years since.
Similarly, niche inflammatory conditions such as \textit{allergic contact dermatitis due to plants} (L23.7) were captured in a single 2020 interval. 
The existence of these orphans underscores the structural fragility of the current data landscape: In the absence of strategic, sustained surveillance pipelines, the field remains dependent on episodic, uncoordinated releases that fail to provide the continuous representation necessary for robust model generalization.

\begin{landscape}

\end{landscape}

\subsection{Detailed results}

\subsubsection{Detailed results on \acrlong*{fst} prediction}

To ensure that the analysis based on skin tones is not confounded by overestimation of the types, we provide confusion matrices in Supplementary Fig.~\ref{fig:ConfusionMatricesFST} for the \gls{fst} prediction on hold-out datasets, \textit{i.e.}, \texttt{DDI}, \texttt{PAD-UFES-20}, and the hold-out test set for the full collection.
Results show that, across datasets, mispredictions mostly occur when confusing boundary types, \textit{i.e.}, types that are similar to each other, such as \gls{fst} V--VI, and less often when confusing higher types for lower ones.
This provides evidence that, although there are mispredictions, they often occur near the ground truth, indicating that the estimates reflect the underlying distributions.

\subsubsection{Dataset distances}

To provide a comprehensive map of the relationships among the 29 aggregated datasets, we computed the pairwise \gls{fd} between all collection pairs and visualized the results as a heatmap in Supplementary Fig.~\ref{fig:DomainShiftHeatmap}. 
This quantitative analysis reveals significant distributional homologies that challenge the assumption of dataset independence. 
We observe a distinct \texttt{ISIC} cluster in which annual challenges (2016--2020) and institutional archives such as \texttt{MSKCC} exhibit minimal pairwise distances, reflecting their shared lineage and patient populations. 
Furthermore, the analysis statistically confirms suspected subset relationships.
The vanishingly small \gls{fd} between \texttt{Fitzpatrick17k} and \texttt{SkinCAP} confirms that the latter functions as a subset of the former, while similar homologies link \texttt{CAN2000} and \texttt{CAN5600}.
Conversely, datasets such as \texttt{PH2} and \texttt{MED-NODE} emerge as distributional outliers, with high average distances to the rest of the field, thereby identifying them as unique sources of variance and critical benchmarks for evaluating model robustness to genuine domain shifts.

\subsection{Additional information}

\subsubsection{Template description and LLM prompt}\label{app:Prompt}

To generate concise captions suitable for multi-modal model training (maximum 77 tokens), we employed an open-source \gls{llm} to compress the template image descriptions while maintaining medical accuracy.
We used the Qwen2.5-7B-Instruct model~\citep{qwen2.5} with the following configuration: Temperature~$=$~0.3, maximum new tokens~$=$~100, top-p sampling~$=$~0.9, running inference in float16 precision on CUDA-enabled hardware.
The compression task was structured using a two-component prompt template.
The system prompt established the task requirements and quality criteria:

\begin{quote}
You are a medical caption compression expert. 
Your task is to rewrite long medical image descriptions into concise captions suitable for multi-modal model training.

\noindent
Requirements:\\
1. Keep all medically relevant information (condition, body location, demographics)\\
2. Remove verbose phrasing and unnecessary words\\
3. Maintain medical accuracy and terminology\\
4. Target maximum length: 77 tokens (typically 60-70 words)\\
5. Use clear, direct language\\

\noindent
Example:\\
Input: ``This total body photograph tile shows a close-up benign skin lesion on the right leg (lower extremity) for a male patient in his 60s.''\\
Output: ``Benign skin lesion on right leg, male, aged 60-69.''
\end{quote}

For each caption, the user prompt template provided the specific compression instruction:

\begin{quote}
Compress this medical image caption while maintaining all medically relevant information:

\noindent
\{original\_caption\}

\noindent
Provide only the compressed caption, no explanation.
\end{quote}

Generated responses were post-processed to extract only the first line of output, removing any explanatory text.
In case of generation failures, the system preserved the original description as a fallback.

\subsubsection{Hyperparameters and configurations}\label{app:hyperparameters}

We detail all model architectures, training configurations, and fusion parameters used to construct the \textsc{SkinMap} embedding space.
The final ensemble comprises eight multi-modal models trained on different data partitions and three self-supervised learning models pretrained on the imaging data alone.

\bmhead{Multi-modal backbone training}
We trained multiple CLIP~\citep{radford2021learning} and SigLIP~\citep{zhai2023sigmoid} models from scratch on the curated image-text pairs.
All models used a fixed input resolution of 224$\times$224 pixels with standard ImageNet normalization~\citep{deng_imagenet_2009}.
Supplementary Table~\ref{tab:clip_training_hyperparameters} summarizes the core training hyperparameters shared across all multi-modal models.

\bmhead{Model ensemble composition}
The ensemble includes models with different architectures (ViT-Large~\citep{dosovitskiy_image_2021} and ViT-Base~\citep{dosovitskiy_image_2021}), loss functions (CLIP~\citep{radford2021learning} vs.\ SigLIP~\citep{zhai2023sigmoid}), and training datasets.
This diversity ensures robustness to model-specific biases and captures complementary aspects of the visual and semantic space.
Supplementary Table~\ref{tab:model_ensemble} details the specific configuration of each model in the ensemble.

\bmhead{Embedding fusion and projection}
Individual embeddings from all 11 encoders were concatenated and projected to a unified 1024-dimensional space using a learnable linear projector.
The projector was trained with a combination of CLIP-style contrastive loss and STRUCTURE regularization~\citep{groger2025limited} to preserve the semantic organization of the pretrained teacher spaces while enabling efficient fusion.
Supplementary Table~\ref{tab:projector_hyperparameters} summarizes the projection training configuration.

\subsection{List of datasets}\label{app:Datasets}
Our analysis integrates 29 publicly available dermatology image datasets, comprising 1,135,082 unique images after deduplication. 
Below, we provide a detailed description of each dataset, including imaging modality, geographic origin, and sample size. 
Dataset counts represent the final deduplicated samples retained in our corpus.

\subsubsection{ISIC archive}
The International Skin Imaging Collaboration (ISIC) archive~\citep{isic2022} constitutes the largest single data source, aggregated across multiple challenge years (2016--2024) and contributing institutions. We treat each challenge year as a separate cohort to preserve temporal metadata.

\begin{itemize}
    \item \textbf{ISIC-2024}~\citep{kurtansky2024slice} (401,059 images): Total body photography dataset from the 2024 challenge, primarily sourced from US institutions. 
    Images employ \gls{tbp} modality for longitudinal lesion monitoring.

    \item \textbf{ISIC-2020}~\citep{rotemberg2021patient} (44,108 images): Dermoscopic images from the 2020 challenge, predominantly contributed by institutions in Australia.

    \item \textbf{ISIC-2019}~\citep{isic2022} (18,946 images): Dermoscopic lesion dataset from the 2019 challenge, sourced primarily from Hospital Clínic de Barcelona, Spain.

    \item \textbf{ISIC-2018}~\citep{codella2019skin} (2,805 images): Dermoscopic images from the 2018 challenge, representing multi-institutional contributions.

    \item \textbf{ISIC-2017}~\citep{codella2018skin} (1,696 images): Dermoscopic dataset from the 2017 challenge focusing on melanoma detection.

    \item \textbf{ISIC-2016}~\citep{gutman2016skin} (1,279 images): Inaugural ISIC challenge dataset with dermoscopic images.

    \item \textbf{ISIC-others}~\citep{isic2022} (1,898 images): Aggregated samples from smaller ISIC collections not associated with annual challenges, predominantly from US institutions.
\end{itemize}

\subsubsection{Clinical photography datasets}
\begin{itemize}
    \item \textbf{Derm1M}~\citep{yan2025derm1m} (412,038 images): Large-scale vision-language dataset aligned with clinical ontologies, aggregated from diverse web-sourced educational materials.

    \item \textbf{Fitzpatrick17k}~\citep{groh_evaluating_2021} (16,574 images): Benchmark dataset for evaluating algorithmic performance across skin tones, compiled from online dermatology atlases with expert-annotated Fitzpatrick skin type labels.

    \item \textbf{SCIN}~\citep{10.1001/jamanetworkopen.2024.46615} (5,033 images): Skin Condition Image Network dataset, crowdsourced from US internet users via Google Search advertisements. Includes self-reported demographics and dermatologist labels across diverse skin tones.

    \item \textbf{PASSION}~\citep{gottfrois2024passion} (4,145 images): Pediatric dermatology dataset from Sub-Saharan Africa, collected 2020--2023 across Madagascar (2,185 images), Guinea (480 images), Malawi (2,150 images), and Tanzania (86 images). Focus on common conditions in pigmented skin: Eczema, fungal infections, scabies, and impetigo.

    \item \textbf{DermaCon-IN}~\citep{madarkar2025dermacon} (5,450 images): Clinical dermatology dataset from India, featuring diverse skin conditions with expert annotations across Fitzpatrick and Monk skin tone scales. Predominantly comprises infectious disorders (2,227 images), inflammatory disorders (1,592 images), and pigmentary disorders (831 images), representing common dermatological presentations in South Asian populations.

    \item \textbf{SD-128}~\citep{leibe_benchmark_2016} (5,619 images): Clinical skin disease classification benchmark derived from the DermQuest online atlas, comprising 128 disease categories.

    \item \textbf{SkinCAP}~\citep{zhou2024skincap} (4,000 images): Mixed-modality dataset combining clinical photography and close-range imaging.

    \item \textbf{PAD-UFES-20}~\citep{pacheco_pad-ufes-20_2020} (2,298 images): Brazilian smartphone-captured lesion dataset from Federal University of Espírito Santo, representing real-world telemedicine imaging conditions.

    \item \textbf{DDI}~\citep{daneshjou_disparities_2022} (656 images): Diverse Dermatology Images dataset from Stanford, retrospectively curated 2010--2020 with pathologically confirmed diagnoses stratified across Fitzpatrick skin types I--VI.

    \item \textbf{MED-NODE}~\citep{giotis2015a} (170 images): Computer-assisted melanoma diagnosis dataset from the Netherlands, employing non-dermoscopic clinical photography.
\end{itemize}

\subsubsection{Dermoscopic datasets}
\begin{itemize}
    \item \textbf{HAM10000}~\citep{tschandl_ham10000_2018} (11,526 images): Multi-source dermoscopic image collection of common pigmented lesions from Medical University of Vienna, Austria, and University of Queensland, Australia.

    \item \textbf{Derm7pt}~\citep{Kawahara2018-7pt} (2,022 images): Seven-point melanoma checklist dataset from Italy, featuring both clinical and dermoscopic modalities with structured diagnostic criteria annotations.

    \item \textbf{PH2}~\citep{mendonca_ph2_2013} (200 images): Dermoscopic benchmark from Portugal with expert-segmented melanocytic lesions.

    \item \textbf{Daffodil}~\citep{khushbu_akter_2024_skin_disease_classification} (9,548 images): Dermoscopic dataset aggregated from web-sourced educational materials.

    \item \textbf{MSKCC}~\citep{mskcc_2025_skin_tone_labeling} (8,984 images): Dermoscopic images from Memorial Sloan Kettering Cancer Center, New York.

    \item \textbf{HIBA}~\citep{ricci2023dataset} (1,635 images): Dermoscopic dataset from Hospital Italiano de Buenos Aires, Argentina.
\end{itemize}

\subsubsection{Educational datasets}
\begin{itemize}
    \item \textbf{PubMed} (126,284 images): Dermatology images extracted from scientific literature in the PubMed Central open-access repository, representing global clinical research.

    \item \textbf{DermNet} (30,324 images): Clinical educational resource from DermNet NZ, New Zealand.

    \item \textbf{DermaCompass} (2,450 images): Clinical educational resource from DermaCompass from Switzerland.

    \item \textbf{Altmeyers} (7,673 images): Clinical dermatology encyclopedia images from Germany.

    \item \textbf{CAN5600} (5,619 images): Web-sourced clinical images from approximately 80 countries.

    \item \textbf{CAN2000} (2,006 images): Subset of web-sourced clinical images from diverse international sources.
\end{itemize}



\end{appendices}


\bibliographystyle{unsrtnat} 
\bibliography{bibliography}

\end{document}